\documentclass[twocolumn]{aastex61}
\usepackage[english]{babel}
\usepackage{gensymb}
\usepackage{textcomp}
\usepackage{ulem}
\usepackage{soul}
\usepackage{amsmath}
\hypersetup{
    colorlinks=true,
    linkcolor=blue,
    filecolor=magenta,      
    urlcolor=blue,
}

\definecolor{ForestGreen}{RGB}{34,139,34}
\definecolor{RoyalBlue}{rgb}{0.25, 0.41, 0.88}

\submitjournal{ApJ}
\received{July 4, 2024}
\accepted{January 6, 2025}

\shorttitle{The DESI view of  the GD-1 Stream and Cocoon}
\shortauthors{Valluri et al.}
\begin{document} 

\title{GD-1 Stellar Stream and Cocoon in the DESI Early Data Release}

\newcommand{\NOIR}{\affiliation{NSF's National Optical-Infrared Astronomy Research Laboratory (NOIRLab)}}
\newcommand{\UMich}{\affiliation{University of Michigan, Department of Astronomy, 1085 S. University Ave, Ann Arbor 48109, MI, USA.}}
\newcommand{\Edinburgh}{\affiliation{University of Edinburgh}}
\newcommand{\Toronto}{\affiliation{University of Toronto}}
\newcommand{\TsingHuaAP}{\affiliation{Institute of Astronomy and Department of Physics, National Tsing Hua University, Hsinchu 30013, Taiwan}}
\newcommand{\TsingHuaIC}{\affiliation{Center for Informatics and Computation in Astronomy, National Tsing Hua University, Hsinchu 30013, Taiwan}}

\author[0000-0002-6257-2341]{M.~Valluri}
\affiliation{Department of Astronomy, University of Michigan, Ann Arbor, MI 48109, USA; email:mvalluri@umich.edu}
\author{P.~Fagrelius}
\affiliation{NSF NOIRLab, 950 N. Cherry Ave., Tucson, AZ 85719, USA}
\author[0000-0003-2644-135X]{S.~E.~Koposov}
\affiliation{Institute for Astronomy, University of Edinburgh, Royal Observatory, Blackford Hill, Edinburgh EH9 3HJ, UK}
\affiliation{Institute of Astronomy, University of Cambridge, Madingley Road, Cambridge CB3 0HA, UK}
\author[0000-0002-9110-6163]{T.~S.~Li}
\affiliation{Department of Astronomy \& Astrophysics, University of Toronto, Toronto, ON M5S 3H4, Canada}
\author[0000-0001-9852-9954]{Oleg~Y.~Gnedin}
\affiliation{Department of Astronomy, University of Michigan, Ann Arbor, MI 48109, USA}
\author[0000-0002-5564-9873]{E.~F.~Bell}
\affiliation{Department of Astronomy, University of Michigan, Ann Arbor, MI 48109, USA}
\author[0000-0002-7667-0081]{R.~G.~Carlberg}
\affiliation{Department of Astronomy \& Astrophysics, University of Toronto, Toronto, ON M5S 3H4, Canada}
\author[0000-0001-8274-158X]{A.~P.~Cooper}
\affiliation{Institute of Astronomy and Department of Physics, National Tsing Hua University, 101 Kuang-Fu Rd. Sec. 2, Hsinchu 30013, Taiwan}

\author{J.~Aguilar}
\affiliation{Lawrence Berkeley National Laboratory, 1 Cyclotron Road, Berkeley, CA 94720, USA}
\author[0000-0001-6098-7247]{S.~Ahlen}
\affiliation{Physics Dept., Boston University, 590 Commonwealth Avenue, Boston, MA 02215, USA}
\author[0000-0002-0084-572X]{C.~Allende~Prieto}
\affiliation{Departamento de Astrof\'{\i}sica, Universidad de La Laguna (ULL), E-38206, La Laguna, Tenerife, Spain}
\affiliation{Instituto de Astrof\'{\i}sica de Canarias, C/ V\'{\i}a L\'{a}ctea, s/n, E-38205 La Laguna, Tenerife, Spain}
\author{V.~Belokurov}
\affiliation{Institute of Astronomy, Madingley Rd, Cambridge CB3 0HA, UK}
\author[0000-0002-0740-1507]{L.~Beraldo e Silva}
\affiliation{Steward Observatory, University of Arizona, 933 N, Cherry Ave, Tucson, AZ 85721, USA}
\author{D.~Brooks}
\affiliation{Department of Physics \& Astronomy, University College London, Gower Street, London, WC1E 6BT, UK}
\author[0000-0002-5689-8791]{A.~Bystr\"om}
\affiliation{Institute for Astronomy, University of Edinburgh, Royal Observatory, Blackford Hill, Edinburgh EH9 3HJ, UK}
\author{T.~Claybaugh}
\affiliation{Lawrence Berkeley National Laboratory, 1 Cyclotron Road, Berkeley, CA 94720, USA}
\author{K.~Dawson}
\affiliation{Department of Physics and Astronomy, The University of Utah, 115 South 1400 East, Salt Lake City, UT 84112, USA}
\author[0000-0002-4928-4003]{A.~Dey}
\affiliation{NSF NOIRLab, 950 N. Cherry Ave., Tucson, AZ 85719, USA}
\author{P.~Doel}
\affiliation{Department of Physics \& Astronomy, University College London, Gower Street, London, WC1E 6BT, UK}
\author[0000-0002-2890-3725]{J.~E.~Forero-Romero}
\affiliation{Departamento de F\'isica, Universidad de los Andes, Cra. 1 No. 18A-10, Edificio Ip, CP 111711, Bogot\'a, Colombia}
\author{E.~Gazta\~naga}
\affiliation{Institut d'Estudis Espacials de Catalunya (IEEC), 08034 Barcelona, Spain}
\affiliation{Institute of Cosmology and Gravitation, University of Portsmouth, Dennis Sciama Building, Portsmouth, PO1 3FX, UK}
\affiliation{Institute of Space Sciences, ICE-CSIC, Campus UAB, Carrer de Can Magrans s/n, 08913 Bellaterra, Barcelona, Spain}
\author[0000-0003-3142-233X]{S.~Gontcho A Gontcho}
\affiliation{Lawrence Berkeley National Laboratory, 1 Cyclotron Road, Berkeley, CA 94720, USA}
\author[0000-0002-6800-5778]{Han,~J}
\affiliation{Center for Astrophysics $|$ Harvard \& Smithsonian, 60 Garden Street, Cambridge, MA 02138, USA}
\author{K.~Honscheid}
\affiliation{Center for Cosmology and AstroParticle Physics, The Ohio State University, 191 West Woodruff Avenue, Columbus, OH 43210, USA}
\affiliation{Department of Physics, The Ohio State University, 191 West Woodruff Avenue, Columbus, OH 43210, USA}
\affiliation{The Ohio State University, Columbus, 43210 OH, USA}
\author[0000-0003-3510-7134]{T.~Kisner}
\affiliation{Lawrence Berkeley National Laboratory, 1 Cyclotron Road, Berkeley, CA 94720, USA}
\author[0000-0001-6356-7424]{A.~Kremin}
\affiliation{Lawrence Berkeley National Laboratory, 1 Cyclotron Road, Berkeley, CA 94720, USA}
\author{A.~Lambert}
\affiliation{Lawrence Berkeley National Laboratory, 1 Cyclotron Road, Berkeley, CA 94720, USA}
\author{M.~Landriau}
\affiliation{Lawrence Berkeley National Laboratory, 1 Cyclotron Road, Berkeley, CA 94720, USA}
\author[0000-0001-7178-8868]{L.~Le~Guillou}
\affiliation{Sorbonne Universit\'{e}, CNRS/IN2P3, Laboratoire de Physique Nucl\'{e}aire et de Hautes Energies (LPNHE), FR-75005 Paris, France}
\author[0000-0003-1887-1018]{M.~E.~Levi}
\affiliation{Lawrence Berkeley National Laboratory, 1 Cyclotron Road, Berkeley, CA 94720, USA}
\author[0000-0002-1769-1640]{A.~de la Macorra}
\affiliation{Instituto de F\'{\i}sica, Universidad Nacional Aut\'{o}noma de M\'{e}xico,  Cd. de M\'{e}xico  C.P. 04510,  M\'{e}xico}
\author[0000-0003-4962-8934]{M.~Manera}
\affiliation{Departament de F\'{i}sica, Serra H\'{u}nter, Universitat Aut\`{o}noma de Barcelona, 08193 Bellaterra (Barcelona), Spain}
\affiliation{Institut de F\'{i}sica dâ€™Altes Energies (IFAE), The Barcelona Institute of Science and Technology, Campus UAB, 08193 Bellaterra Barcelona, Spain}
\author[0000-0002-4279-4182]{P.~Martini}
\affiliation{Center for Cosmology and AstroParticle Physics, The Ohio State University, 191 West Woodruff Avenue, Columbus, OH 43210, USA}
\affiliation{Department of Astronomy, The Ohio State University, 4055 McPherson Laboratory, 140 W 18th Avenue, Columbus, OH 43210, USA}
\affiliation{The Ohio State University, Columbus, 43210 OH, USA}
\author[0000-0003-0105-9576]{G.~E.~Medina}
\affiliation{Department of Astronomy \& Astrophysics, University of Toronto, Toronto, ON M5S 3H4, Canada}
\author[0000-0002-1125-7384]{A.~Meisner}
\affiliation{NSF NOIRLab, 950 N. Cherry Ave., Tucson, AZ 85719, USA}
\author{R.~Miquel}
\affiliation{Instituci\'{o} Catalana de Recerca i Estudis Avan\c{c}ats, Passeig de Llu\'{\i}s Companys, 23, 08010 Barcelona, Spain}
\affiliation{Institut de F\'{i}sica dâ€™Altes Energies (IFAE), The Barcelona Institute of Science and Technology, Campus UAB, 08193 Bellaterra Barcelona, Spain}
\author[0000-0002-2733-4559]{J.~Moustakas}
\affiliation{Department of Physics and Astronomy, Siena College, 515 Loudon Road, Loudonville, NY 12211, USA}
\author{A.~D.~Myers}
\affiliation{Department of Physics \& Astronomy, University  of Wyoming, 1000 E. University, Dept.~3905, Laramie, WY 82071, USA}
\author[0000-0002-5758-150X]{J.~Najita}
\affiliation{NSF NOIRLab, 950 N. Cherry Ave., Tucson, AZ 85719, USA}
\author{C.~Poppett}
\affiliation{Space Sciences Laboratory, University of California, Berkeley, 7 Gauss Way, Berkeley, CA  94720, USA}
\affiliation{University of California, Berkeley, 110 Sproul Hall \#5800 Berkeley, CA 94720, USA}
\author[0000-0001-7145-8674]{F.~Prada}
\affiliation{Instituto de Astrof\'{i}sica de Andaluc\'{i}a (CSIC), Glorieta de la Astronom\'{i}a, s/n, E-18008 Granada, Spain}
\author[0000-0001-5589-7116]{M.~Rezaie}
\affiliation{Department of Physics, Kansas State University, 116 Cardwell Hall, Manhattan, KS 66506, USA}
\author{G.~Rossi}
\affiliation{Department of Physics and Astronomy, Sejong University, Seoul, 143-747, Korea}
\author[0000-0001-5805-5766]{A.~H.~Riley}
\affiliation{Institute for Computational Cosmology, Department of Physics, Durham University, South Road, Durham DH1 3LE, UK}
\author[0000-0002-9646-8198]{E.~Sanchez}
\affiliation{CIEMAT, Avenida Complutense 40, E-28040 Madrid, Spain}
\author{D.~Schlegel}
\affiliation{Lawrence Berkeley National Laboratory, 1 Cyclotron Road, Berkeley, CA 94720, USA}
\author{M.~Schubnell}
\affiliation{Department of Physics, University of Michigan, Ann Arbor, MI 48109, USA}
\affiliation{University of Michigan, Ann Arbor, MI 48109, USA}
\author{D.~Sprayberry}
\affiliation{NSF NOIRLab, 950 N. Cherry Ave., Tucson, AZ 85719, USA}
\author[0000-0003-1704-0781]{G.~Tarl\'{e}}
\affiliation{University of Michigan, Ann Arbor, MI 48109, USA}
\author[0000-0002-2468-5521]{G.~Thomas}
\affiliation{Instituto de Astrof\'{\i}sica de Canarias, C/ V\'{\i}a L\'{a}ctea, s/n, E-38205 La Laguna, Tenerife, Spain
Departamento de Astrof\'{\i}sica, Universidad de La Laguna (ULL), E-38206, La Laguna, Tenerife, Spain
}
\author{B.~A.~Weaver}
\affiliation{NSF NOIRLab, 950 N. Cherry Ave., Tucson, AZ 85719, USA}
\author[0000-0003-2229-011X]{R.~H.~Wechsler}
\affiliation{Kavli Institute for Particle Astrophysics and Cosmology, Stanford University, Menlo Park, CA 94305, USA}
\affiliation{Physics Department, Stanford University, Stanford, CA 93405, USA}
\affiliation{SLAC National Accelerator Laboratory, Menlo Park, CA 94305, USA}
\author[0000-0001-5381-4372]{R.~Zhou}
\affiliation{Lawrence Berkeley National Laboratory, 1 Cyclotron Road, Berkeley, CA 94720, USA}
\author[0000-0002-6684-3997]{H.~Zou}
\affiliation{National Astronomical Observatories, Chinese Academy of Sciences, A20 Datun Rd., Chaoyang District, Beijing, 100012, P.R. China}

\def\vector#1{\mbox{\boldmath $#1$}}
\newcommand{\AU}{\ensuremath{\,\mathrm{AU}}}
\newcommand{\pc}{\ensuremath{\,\mathrm{pc}}}
\newcommand{\kpc}{\ensuremath{\,\mathrm{kpc}}}
\newcommand{\Myr}{\ensuremath{\,\mathrm{Myr}}}
\newcommand{\Gyr}{\ensuremath{\,\mathrm{Gyr}}}
\newcommand{\kms}{\ensuremath{\,\mathrm{km\ s}^{-1}}}
\newcommand{\kmskpc}{\ensuremath{\,\mathrm{{km\ s}^{-1}\ {kpc}^{-1}}}}
\newcommand{\msun}{\ensuremath{\,M_\odot}}
\newcommand{\mas}{\ensuremath{\,\mathrm{mas}}}
\newcommand{\masyr}{\ensuremath{\,\mathrm{mas\ yr}^{-1}}}
\newcommand{\microas}{\ensuremath{\,\mu\mathrm{as}}}
\newcommand{\microasyr}{\ensuremath{\,\mu\mathrm{as\ yr}^{-1}}}
\newcommand{\Kelvin}{\ensuremath{\,\mathrm{K}}}
\newcommand{\vlos}{v_{\ensuremath{\mathrm{los}}}}

\newcommand{\rapo}{r_\ensuremath{\mathrm{apo}}}
\newcommand{\rperi}{r_\ensuremath{\mathrm{peri}}}
\newcommand{\zmax}{z_\ensuremath{\mathrm{max}}}
\newcommand{\flag}[1]{\texttt{\lowercase{#1}}}
\newcommand{\agama}{\texttt{Agama}}
\newcommand{\emcee}{\flag{emcee}}
\newcommand{\Gaia}{\textit{Gaia}}
\newcommand{\nbody}{\textit{N}-body}

\newcommand{\eq}[1]{\begin{align}#1\end{align}}

\begin{abstract}
We present { $\sim 115$ } new spectroscopically identified members of the GD-1 tidal stream observed with the 5000-fiber Dark Energy Spectroscopic Instrument (DESI). We confirm the existence of a ``cocoon'' which is a broad (FWHM$\sim 2.932^\circ \sim 460$pc) and kinematically hot (velocity dispersion, $\sigma \sim 5-8$\kms) component that surrounds a narrower (FWHM$\sim 0.353^\circ \sim 55$pc) and colder  {($\sigma =3.09\pm0.76$\kms)} thin stream component (based on a median per star velocity precision of 2.7~\kms). The cocoon extends over at least a  $30^\circ$ segment of the stream observed by DESI. The thin and cocoon components have similar mean  values of [Fe/H]: $-2.54\pm 0.04$~dex and $-2.47\pm 0.06$~dex suggestive of a common origin.  The data are consistent with the following scenarios for the origin of the cocoon.  The progenitor of the GD-1 stream was an accreted globular cluster (GC) and: (a) the cocoon was produced by pre-accretion tidal stripping of the GC while it was still inside its parent dwarf galaxy; (b) the cocoon is debris from the parent dwarf galaxy; (c) an initially thin GC tidal stream was heated by impacts from dark subhalos in the Milky Way; (d) an initially thin GC stream was heated by a massive Sagittarius dwarf galaxy; or a combination of some these.  Future DESI spectroscopy and detailed modeling may enable us to distinguish between these possible origins.
\end{abstract}

\keywords{Unified Astronomy Thesaurus concepts:
Atomic spectroscopy (2099), Spectroscopy (1558), Dark matter (353), Stellar streams (2166), Milky Way dynamics (1051); Milky Way dark matter halo (1049)}

\section{Introduction \label{sec:intro}}

Stellar streams are a stunning example of the ongoing hierarchical assembly of the Milky Way via the gravitational interaction between our Galaxy and its satellites. They are the result of the disruptive tidal forces of the Galaxy on its companions such as globular clusters and dwarf galaxies. If the progenitor of a tidal stream is of low enough mass, the tidal debris that is liberated follows a path that is close to the orbit of its progenitor.

Long tidal streams that extend over tens or even hundreds of degrees on the sky are of particular interest since they can be used to probe  the global mass density profile, total mass and 3D shape of the Milky Way's dark matter halo \citep[e.g.,][]{helmi_2004, johnston_etal_2005, law_etal_2009, Law2010, koposov2010, Malhan_2019_shape,tango2021}. A stellar stream is the result of tidal fields removing stars from the outskirts of a progenitor at a velocity dispersion close to the cluster velocity dispersion, with the tide at the stream pericenter creating an outward ordered angular momentum distribution that leads to orbital shearing and a wider orbital spread at apocenter. Since globular clusters (GCs) have small internal velocity dispersions ($\lesssim 2$\kms), unperturbed tidal streams produced by GCs are both thin and kinematically cold. Therefore velocity dispersion variations,  density perturbations and gaps along a GC stream might hold clues to past fly-by interactions and impact from dark matter subhalos which can produce gaps and heating which results in broadening the streams over time. If the streams do not pass through the Galactic disk they may be especially good antennae for detecting such dark subhalos.  Consequently observations of both the spatial and velocity substructure in tidal streams have elicited great interest as potential probes of dark matter substructure \citep[e.g.,][]{Johnston_2002,Ibata_etal_2002, siegal-gaskins_valluri_2008,Carlberg2012,Carlberg2013,Erkal_2015a,Bonaca_2019}.  

The GD-1 stellar stream, first identified by \cite{grillmair_2006}, likely originated as a GC \citep{koposov2010}. It is located in the Galactic halo, and being both thin and very long (spanning at least 100$^{\circ}$), is a good candidate for exploring both the global distribution of dark matter in the inner Milky Way \citep[e.g.,][]{koposov2010,Bowden2015,Malhan_2019_shape} and for detecting dark subhalos from the analysis of gaps and velocity dispersion variations. The progenitor remains undetected, possibly having been completely disrupted \citep{deBoer2018, Malhan2018, Price-Whelan_2018}. GD-1 is of great current interest since Gaia DR2 data led to the detection of density variations, gaps and off-stream structures e.g., a `spur' a `blob' and `wiggles' \citep[][]{Price-Whelan_2018,deBoer2020} which have been claimed as evidence of possible impacts by dark matter subhalos \citep{Bonaca_2019}. Since the stream does not pass through the inner disk region (it has a perigalactic (apogalactic) distance of $\sim 14 ~ (26)$~kpc, and inclination of 39$^\circ$ to the disk plane, \cite{koposov2010}) it is unlikely to have been heated by encounters with giant molecular clouds in the inner disk. It has also been shown that GD-1's orbit has a very low probability of impact with other (intact) GCs  \citep{Doke_Hattori_2022}. 

An alternative explanation for the gaps and density variations in the GD-1 stream has been proposed: regularly spaced gaps have been predicted to arise naturally due to epicyclic motions of stars along the stream \citep{Kupper_2012,Kupper_2015}. From an analysis of the spectrum of stream gaps and after accounting for artifacts in the Gaia DR2 scanning law and projection effects, \citet{IbataGD1gaps2020} conclude that the density profile of the stream shows periodic gaps separated by $\sim 2.64\pm 0.18$kpc and that such a regularly spaced gap distribution can be produced via epicyclic motions in a smooth Galactic potential. However since the regularly spaced gaps arising from internal dynamics are not predicted to be  associated with increased velocity dispersions, obtaining kinematic data is useful for distinguishing this process from the effects of perturbations.

\citet{Malhan_2019_cocoon} used Gaia DR2 data to show that the GD-1 stellar stream  possesses a secondary diffuse and extended stellar component ($\sim 100$~pc wide) that surrounds the previously identified thinner component of the stream. They named this broader component the ``cocoon'' and showed that it was detected at $> 5\sigma$ confidence level. Similar broad and extended morphological and kinematic features have also been found in several other streams \citep[e.g., Jhelum,  Phlegethon,][]{Malhan_2020,Malhan_2022_pm, Ibata_2023,Awad_2024}. Simulated GC streams that form in dwarf galaxies  and are then accreted into Milky Way like halos show complex morphological features (like gaps, ``spurs'', ``blobs'', ``cocoons''). Such accreted GC streams have complex morphologies in both  cosmological simulations \citep{Carlberg_2018,Carlberg_alger_2023} and  controlled simulations \citep{Malhan_2020}. These complex morphologies are  produced via the tidal interactions between the GCs and the central dark matter distributions in the GC's parent dwarf galaxy prior to their infall and subsequent accretion onto a massive host galaxy like the Milky Way. Furthermore it has been shown that the phase space structure of accreted GC streams could be used to obtain constraints on the dark matter distribution in the parent dwarf galaxy \citep{Malhan_2020,Malhan_2022_pm}, making the phase space structure of the cocoon important to determine. 

Recently \citep{Dillamore_etal_2022} showed that the GD-1 stream is extremely sensitive to perturbations from the Sagittarius dwarf galaxy (Sgr) especially if it was quite massive ($\gtrsim 4\times 10^{10}$\msun) since the orbit of the GD-1 stream intersects with that of Sgr. They showed that a wide range of outcomes is possible with various types of substructures produced in and around the GD-1 stream (see their Fig 9). Interactions with the Sgr dwarf change the energy distribution of the GD-1 particles making it substantially broader; also the streams density profile is altered (affecting the surface brightness distribution along the stream) and its length is modified. While many of the induced perturbations are difficult to distinguish from the cocoon formation scenario above, the change in length is particularly interesting as this is a unique prediction of the heavy Sgr hypothesis.

Since impacts from dark matter sub-halos which cross a stream also cause stream heating (in addition to producing shortlived gaps and off-stream features like `spurs'), it is particularly important to measure the velocity dispersions of robustly identified stream candidates. Thus the phase space structure of the GD-1 stream  has the potential to improve our understanding of the nature of dark matter both within the Milky Way and in dwarf satellites that were accreted by our Galaxy.

While Gaia photometry and proper-motions led to the detection of an extension to the stream as well as these various complex morphological features, there are fewer than 100 spectroscopically and chemically confirmed members from previous surveys with high radial velocity uncertainty of $\sim 1-2$\kms\ \citep{Bonaca_2020,Gialluca_2021,IbataGD1gaps2020}. An additional $\sim 200$ spectroscopically confirmed members are available from the SDSS and LAMOST surveys but have lower radial velocity precision ($\sim 5-20$\kms).  In this work we present 115 new spectroscopically confirmed members of the GD-1 stream obtained with the Dark Energy Spectroscopic Instrument (DESI)  with $\sim 1-10$\kms~velocity precision (median precision 2.7\kms) and $\sim 0.2$~dex metallicity precision. We also improve on  17 previous radial velocity measurements.  

Starting in December 2020, DESI carried out an approximately 6 month long ``Survey Validation'' (SV) phase. The data from SV were used to validate scientific requirements, specifically the target selection algorithms \citep{DESI_Instr_Overview2022}. During the SV period, while testing and streamlining targeting algorithms and analysis pipelines, there was also an opportunity to observe specific targets. 
In this period, DESI took several observations in the GD-1 stream region. Since these data were obtained primarily as part of the SV program, the target selection criteria and exposure times were slightly different from the main survey. After the completion of SV, there have continued to be observations in the GD-1 stream area. 

In this paper we introduce the spectroscopically identified stars associated with the GD-1 tidal stream from the DESI SV observations, which was recently released to the public \citep{desi_edr, koposov_edr_2024}. 
Additional data releases are planned at regular intervals. In section \ref{sec:data}, we briefly  describe the DESI Survey Validation data set in the region overlapping GD-1. In section~\ref{sec:membership} we discuss how we identify GD-1 stream members from the DESI targets using cuts in the stellar color-magnitude diagram, Gaia parallax, Gaia proper-motions as well as radial velocity and metallicity [Fe/H]. In section \ref{sec:cocoon} we discuss the phase space distribution of stars in the stream, focusing on the identification and kinematics of the broader  region referred to as the ``cocoon''. Finally, in Section~\ref{sec:discussion} we summarize our findings and discuss the implications of these results. We also give a preview of what to expect from future DESI data releases on the GD-1 stream region.

\section{DESI observations of the GD-1 region}
\label{sec:data}

\begin{figure*}[th]
    \includegraphics[width=\textwidth]{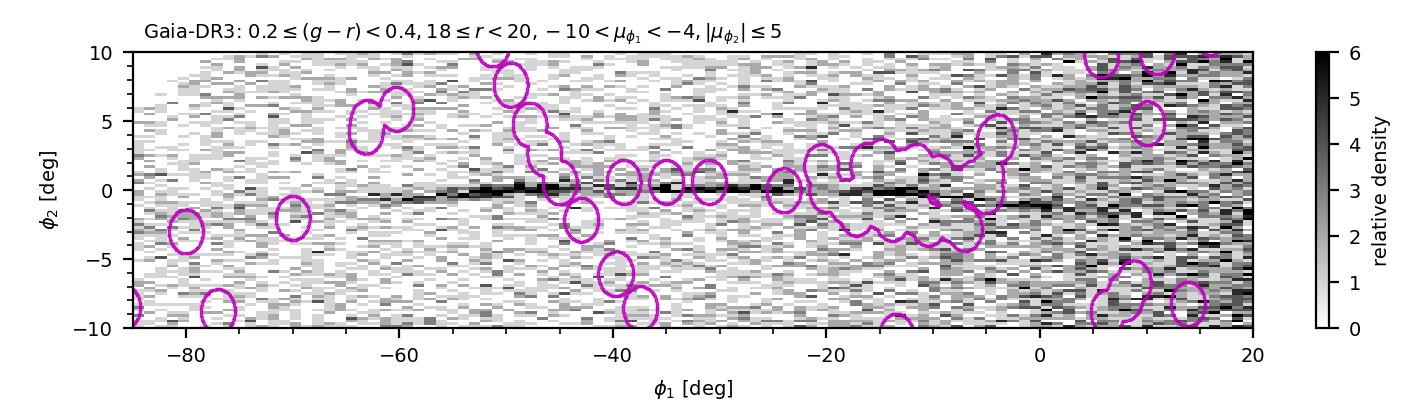}
\caption{GD-1 stream in great-circle coordinates ($\phi_1, \phi_2$) with the stream along $\phi_1$. Grey scale shows 2D histogram  of surface density of Gaia DR3 stars with a broad color-magnitude, and a proper motion selection as indicated in the legend. The GD-1 stream is clearly visible as an over-density at roughly  $\phi_2=0$. Magenta curves show the DESI tiles observed during commissioning and Survey Validation.}
\label{fig:gd1_region}
\end{figure*}

The Dark Energy Spectroscopic Instrument (DESI)  is a robotic, fiber-fed, highly multiplexed spectrograph installed on the Mayall 4-meter telescope at Kitt Peak National Observatory \citep{DESI_Instr_Overview2022}. A central feature of the instrument is the use of 5,000 robotic positioners that move individual optical fibers to pre-identified targets over a $\sim 3^\circ$ field of view \citep{desi16a, Silber:2023aa, miller_2023_desi}. The DESI survey is  a five-year survey of about 14,000 square degrees to obtain spectra for approximately 40 million galaxies and quasars and over 10 million  Milky Way stars \citep{desi16a}.

DESI  was installed in 2019 and commissioning and survey validation extended through May 2021 \citep{desi_sv}. Since May 2021, the 5-year DESI survey has been underway and has obtained an unprecedented number of galaxy, quasar and stellar spectra.  The primary scientific goal of the survey is to use spectroscopic redshifts of galaxies and quasars to obtain the most precise constraints on the expansion history of the universe ever obtained using the baryon acoustic oscillation and other methods (e.g., redshift space distortions) \citep{desi16a, levi13}.

Over the first 5-years  of the DESI Milky Way Survey  \citep{allende20,Cooper2023overview} it is expected that spectra of $\simeq 7$ million unique Milky Way stars at Galactic latitudes $|b| > 20^\circ$ will be obtained during the main survey and millions more will be obtained during poor sky conditions (the ``back up program''). The DESI-Milky Way Survey (MWS) uses a simple and inclusive target selection scheme focused on the thick disk and stellar halo. All targets are selected from the Gaia catalog (originally DR2, but subsequently DR3). Targets fall into three categories: MAIN-BLUE, MAIN-RED, and MAIN-BROAD, within the magnitude range $16<r<19$. MAIN-BLUE prioritizes all Gaia point sources in this magnitude range with blue optical colors  ($g-r < 0.7$) in the DESI Legacy Imaging Surveys \citep{dey19}. The MAIN-RED selection applies Gaia proper-motion and parallax criteria to sources with redder colors $(g-r > 0.7)$ to increase the probability of observing distant halo giants. Sources with $g-r>0.7$ that do not meet the astrometric criteria are targeted at lower priority in the MAIN-BROAD category \citep{Cooper2023overview}.  DESI-MWS is expected to achieve an average spectroscopic completeness of $\sim 28$\% over the various MAIN target classes. 

The DESI-MWS is executed during bright-sky conditions (when high-redshift galaxy observations are inefficient) and shares the focal plane with a low-redshift Bright Galaxy Survey \citep[BGS;][]{Hahn_2023BGS}, although BGS targets have higher priority for fiber assignment than MWS stars. 
The DESI spectroscopic data presented in this paper were taken during the on-sky DESI SV from 2020/12/14 to 2021/5/14; the corresponding data (both spectra and the catalogs used in this paper) are now publically available under DESI's Early Data Release (EDR) \url{https://data.desi.lbl.gov/doc/releases/edr/vac/mws/}. During the SV, the DESI Collaboration carried out observations to evaluate the targeting pipelines, instrument and observatory readiness in three separate subprograms SV1, SV2 and SV3.  The MWS target selection for SV was previously described in \citet{allende20}. Other DESI SV programs are discussed in \citet{desi_sv}. Each fiber configuration in a given field is referred to as a ``tile''. Some of these tiles had a larger percentage of fibers dedicated to Milky Way stars than expected for the main survey and in some cases went to fainter magnitudes (for the SV, stars in the range $19 < r < 20$ were also included at lower priority.) As discussed in Section~7 of \citet{Cooper2023overview},  several fields in the  GD-1 region were observed during SV1 but without any special prioritization of potential GD-1 targets.  
In addition, 22 tiles were observed during SV2, also without any special prioritization for GD-1 stars. In total 45 tiles in the GD-1 region were observed during Survey Validation. For more details on the observing conditions and various programs for SV, readers are directed to the MWS EDR Value Added Catalogue paper \citep{koposov_edr_2024}.

Figure~\ref{fig:gd1_region} shows the GD-1 stream region in a coordinate system in which $\phi_1$ is the angle on the sky along the stream and $\phi_2$ is the angle perpendicular to $\phi_1$. The transformation from Gaia DR3 sky positions ($\alpha$, $\delta$) and proper motions ($\mu_{\alpha}^*, \mu_{\delta}$)
to this natural GD-1 frame  was carried out using the coordinate transformation described in \citet{koposov2010}\footnote{The pole of the great circle frame fitting the GD-1 stream has (RA, Dec)= (34$^\circ$.5987, 29$^\circ$.7331) in 2000 coordinates.}, which also yields proper motion in [$\phi_{1}, \phi_{2}$], hereafter referred to as $\mu_{\phi_1}$ and $\mu_{\phi_2}$\footnote{Hereafter we drop the $^*$ and simply use $\mu_{\alpha}$ and $\mu_{\phi_1}$ to denote $\mu_{\alpha}\cos(\delta)$ and   $\mu_{\phi_1}\cos(\phi_2)$}  respectively. Most targets were also identified in the DESI Legacy Survey (DECaLs) \citep{dey19}, from which  the $g$ and $r$-band magnitudes for these stars were measured.  The Legacy Survey magnitudes were de-reddened using dust maps \citep{Schlafly_Finkbeiner_Schlegel_2010} updated for the Legacy Survey and available on the Legacy Survey website \url{https://www.legacysurvey.org}.
These data were cross-matched with the Gaia DR3 catalog for parallax and proper motion measurements \citep{Gaia_DR3_2022}. The Gaia DR3 proper motions were corrected for solar reflex motion using Astropy. We used Astropy's default parameters for {\tt galactocentric\_frame\_defaults.set(`v4.0')} \citep{Reid_Brunthaler_2004,Gravity_2018,DrimmelPoggio_2018,Bennett_Bovy_2019}. Distances to individual stars were estimated using the distance modulus in Equation~\ref{eq:dist_mod} below (also see Appendix~\ref{sec:distance}).  Gaia parallaxes were zero-point corrected following the prescriptions in \citep{Lindegren_2021_zpt}. 

The 2D histogram in Figure~\ref{fig:gd1_region} shows the surface density of Gaia DR3 stars with DECaLS $0.2 \leq (g-r)< 0.4$ and  $18\leq r <20$,  $-10$mas~yr$^{-1}<\mu_{\phi_1}<-4$mas~yr$^{-1}$ and $|\mu_{\phi_2}|<-5$mas~yr$^{-1}$. These selections are similar to and  based on previous works \citep[e.g.,][]{Price-Whelan_2018}. The boundaries of the $45$ DESI tiles in this region that were observed during SV are shown by magenta curves. The direction of motion of the stream is from positive $\phi_1$ to negative $\phi_1$.

\begin{figure*}[t!]

\includegraphics[trim = 0 0 0 0, clip]{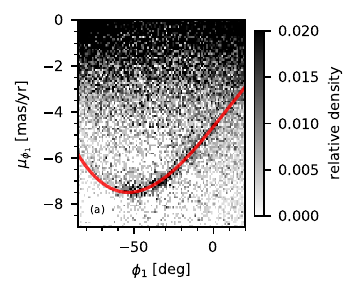}
\includegraphics[trim = 0 0 0 0, clip]{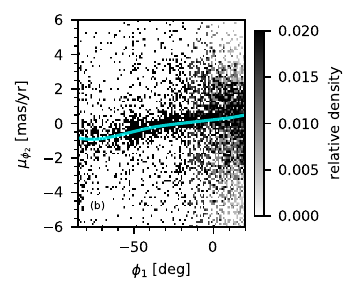}
\includegraphics[trim = 0 0 0 0, clip]{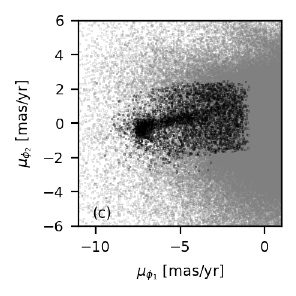}
\includegraphics{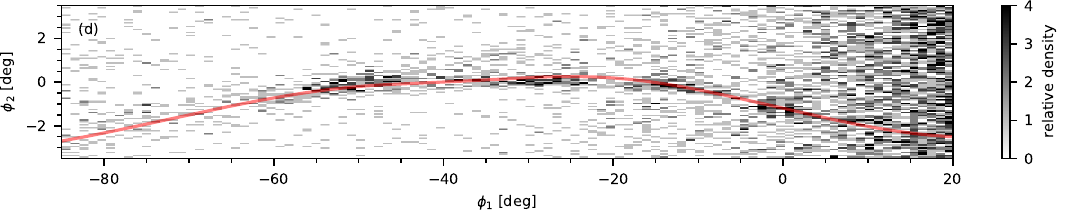}
 \caption{Gaia-DR3 stars in the GD-1 region defined in the frame of reference of the stream [$|\phi_{2}| < 3^\circ$, $-85^\circ < \phi_{1} < 20^\circ$] with a broad color-magnitude cut of $0.2 \leq g-r < 0.4$ and $18\leq r<20$. (a) 2D histogram of $\mu_{\phi_1}$ vs. $\phi_1$ with a spline  (red curve) highlighting the GD-1 stream track. (b) 2D histogram of $\mu_{\phi_2}$ vs. $\phi_1$ with $\mu_{\phi_1}$ within 2mas/yr from red stream track in panel-(a). Cyan curve shows spline fit to stream. (c) scatter plot of $\mu_{\phi_2}$ vs. $\mu_{\phi_1}$. Grey dots are all stars in  panel-(b). Black points are stars  with separation from red and cyan splines with $\delta\mu_{\phi_1}<2$mas/yr and $\delta\mu_{\phi_2}<2$mas/yr and parallax and broad color-magnitude cuts as described in the text.
    (d) Selection in  panel-(c) shown as a 2-D histogram in $\phi_1-\phi_2$ (relative density in arbitrary units) with a spline representing stream track shown in red.}
\label{fig:mu1mu2}
\end{figure*}

\begin{figure}[t!]
\includegraphics[trim= 15 15 0 0, clip,width=\columnwidth]{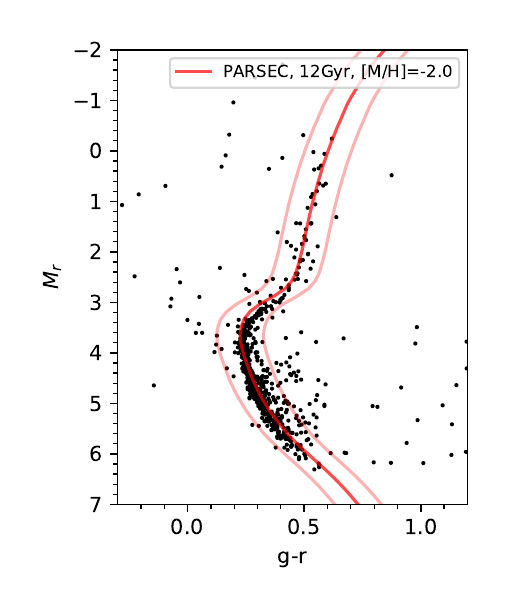}
\caption{Color absolute magnitude diagram with the dereddened magnitudes from DECaLS. For this plot stars were selected with $|\delta\phi_2|<0.2$deg from the stream track and both $\mu_{\phi_1}$ and $\mu_{\phi_2}$ within 0.5~mas/yr from the respective splines in Figure~\ref{fig:mu1mu2}. This selection is only used to define the color-magnitude selection. We use the distance modulus versus $\phi_1$ relation in Eqn.~\ref{eq:dist_mod} to correct for distance gradient along the stream. Overlaid is an isochrone from the PARSEC Isochrone database with a metallicity [M/H] = -2.0~dex and age = 12~Gyr (red curve). The light red curves are at $\Delta|g-r|=0.1$ from the red curve.}
\label{fig:cmd}
\end{figure}

The stellar spectra obtained by the DESI spectrograph are processed with the RVS pipeline, based on {RVSpec} \citep{RVSpec}, to measure each star's line-of-sight (radial) velocity and stellar atmospheric parameters  {($T_\mathrm{eff}, \log(g),$ [Fe/H], [$\alpha$/Fe]). Using the radial velocities from RVS, the spectra are independently processed by the SP pipeline which is based on FERRE \citep{Allende_Prieto_2006}\footnote{
https://github.com/callendeprieto/ferre and supporting python code {\tt piferre} https://github.com/callendeprieto/piferre}. In addition to the 4 atmospheric parameters, above the SP pipeline measures the abundances of Mg, C, Ca.  

The RVS and SP pipeline outputs have been calibrated against the APOGEE measurements. The DESI EDR dataset contains spectra for 507  stars that are also in the APOGEE DR17 catalog    \citep{koposov_edr_2024}. A  detailed study of the random and systematic errors on DESI stellar atmospheric parameters compared with the corresponding APOGEE measurements is presented in section 5.1 of that paper. 
Their analysis shows that the median DESI metallicity has a median offset [Fe/H]$_\mathrm{DESI}-$[Fe/H]$_\mathrm{APOGEE}=0.06^{+0.13}_{-0.18}$~dex for the RVS pipeline and  a median offset $0.03^{+0.18}_{-0.16}$~dex for the pipeline.  For further details of the pipelines and their validation see \citet{Cooper2023overview}  and \citet{koposov_edr_2024}. Since the DESI pipelines are still being improved and the systematic error likely depends on the stellar population, we do not apply any corrections for systematic uncertainties in [Fe/H] in this work.}

 {Despite the good performance of both RVS and SP pipelines on the 507 APOGEE stars however,  \citet{koposov_edr_2024} point out that the SP pipeline output is extremely sensitive to the S/N of the spectra, $T_\mathrm{eff}$, and discontinuities in the stellar templates, and therefore several quality cuts are recommended to obtain robust results (see Appendix A of \citet{koposov_edr_2024} for details). We find that less than 50\% of the stars in the GD-1 sample pass the quality cuts needed for robust SP abundances but all pass the RVS cuts. Therefore for the rest of this paper we restrict ourselves to using the output of the RVS pipeline.}  

\section{GD-1 Stream Membership }
\label{sec:membership}

\subsection{Initial selection}
\label{sec:pm}

Our search for GD-1 stream members begins by restricting to the region $|\phi_{2}| < 3^\circ$, $-85^\circ < \phi_{1} < 20^\circ$ (galactic coordinates: $210^\circ < l <90^\circ$, $20^\circ<b<70^\circ$). We also apply the same broad cut in color-magnitude space and proper-motion selections as  used in Figure~\ref{fig:gd1_region}. These selections expand slightly on the limits set by  \cite{Price-Whelan_2018} and \cite{Malhan_2019_shape}. 

The distance to the GD-1 stream is known to vary with $\phi_1$ \citep{koposov2010}. We assume a distance modulus $D$ given by, 
\begin{eqnarray}
D = 18.82 + [(\phi_1 + 50) / 64]^2 - 4.45 ,\label{eq:dist_mod}
\end{eqnarray}
 which is similar to the  function estimated by \citet{koposov2010}. 
This relation is a simple quadratic fit to the subgiant branch stars in GD-1. More details are provided in Appendix~\ref{sec:distance} where
we also compare this relation with the distance relations proposed by several other authors \citep{Price-Whelan_2018, deBoer2018,LiYannyWu2018}. The functions defined by \citet{deBoer2018} and \citet{LiYannyWu2018} differ only slightly from the distance-$\phi_1$ relation in Equation~\ref{eq:dist_mod} (see Figure~\ref{fig:distance_gradient} in Appendix~\ref{sec:distance}). We use  equation~\ref{eq:dist_mod} in the rest of the paper. 

We impose a Gaia parallax cut such that 
 \begin{eqnarray}
\delta\varpi & =  |1/d(\phi_1) -\varpi|,\\
\delta\varpi & <\varpi_{sys} + 2*\varpi_{err},
\end{eqnarray}
where $d(\phi_1)$ is the distance to the stream at a given value $\phi_1$ computed assuming the distance modulus in Equation~\ref{eq:dist_mod}, $\varpi_{err}$ is the uncertainty on the Gaia parallax for each star, and $\varpi_{sys}=0.05$mas is the systematic error on the parallax \citep{Lindegren2021}.

Figure \ref{fig:mu1mu2} shows three projections of Gaia stars in the GD-1 stream region: the stream is clearly offset from  the background in $\mu_{\phi_1}$ vs. $\phi_1$ (panel-a) and $\mu_{\phi_2}$ vs. $\phi_1$ (panel-b).  We fit  splines to the stream tracks that emerge in  panels a and b (shown by red and cyan curves).  {With the above selections  we  compute $\delta\mu_{\phi_1}$ and $\delta\mu_{\phi_2}$ relative to the  proper motion stream tracks. The densest part of the 2D distribution in $\delta\mu_{\phi_1}$ and $\delta\mu_{\phi_2}$ is reasonably well fitted by a 2D circular Gaussian of standard deviation 0.5~mas/yr. We conservatively select stars within 5 times this standard deviation i.e. all stars within $\pm$2~mas/yr around each proper motion spline} to obtain the selection  shown by black points in Figure~\ref{fig:mu1mu2}~(c) which is similar to the polygon defined in \cite{Price-Whelan_2018}.  Figure~\ref{fig:mu1mu2}~(d) shows the stream in $\phi_1-\phi_2$ coordinates over-plotted with a cubic spline through the highest density regions of the stream shown in red. The values at the knots as a function of $\phi_1$ for these three splines as well as for the velocity spline in Figure~\ref{fig:rv_phi1_all}~(a) (to be discussed in Section~\ref{sec:radial_vel}) are given in Table~1. 

\begin{table}
\centering
\begin{tabular}{rrrrr}
\multicolumn{5}{r}{{\bf Table~1: Knots for four stream-track splines}}\\
\hline 
$\phi_1$ &
$\mu_{\phi_1}$ &
 $\mu_{\phi_2}$ & 
${\phi_2}$ &
 $V_{GSR}$ \\
 
[$^\circ$]& [mas/yr] & [mas/yr] & [$^\circ$] & [km/s] \\
\hline
-90.0 & -3.00 & -4.32 & -0.70 & 75.49\\
-70.0 & -1.50 & -6.92 & -0.88 & 60.78 \\
-60.0 & -0.72 & -7.46 & -0.70 & 29.15 \\
-46.0 & -0.10 & -7.54 & -0.37 & -21.31 \\
-37.6 & 0.05 & -7.29 & -0.21 & -46.37 \\
-29.2 & 0.22 & -6.85 & -0.09 & -66.46 \\
-20.8 & 0.21 & -6.28 & 0.00 & -93.05 \\
-12.4 & -0.18 & -5.61 & 0.08 & -113.76 \\
-4.0 & -0.84 & -4.90 & 0.16 & -133.84 \\
10.0 & -2.03 & -3.70 & 0.32 & -189.77 \\
  \hline \\
\end{tabular}
\label{tab:splines1234}
\end{table}

\subsection{Color-Magnitude Diagram selection  \label{sec:CMD}}

Given the above initial cuts as an input, we show that the data are well identified in a color-magnitude diagram (CMD) as shown in Figure \ref{fig:cmd}.  The stars are well matched to an isochrone of 12~Gyr with [M/H]~$ = -2.0 $ from the PARSEC isochrone database \citep{PARSEC_iso_2012,Tang2014, Marigo2017, Pastorelli2019, Pastorelli2020}\footnote{\url{https://people.sissa.it/~sbressan/parsec.html}}. 

Other authors have used similar isochrones selections \citep[e.g.,][]{IbataGD1gaps2020}.
We accepted stars along the PARSEC isochrone if they lie within $\delta(g-r) \pm 0.1$ (shown by faint red curves). 

\subsection{Radial Velocity Distribution \label{sec:radial_vel}}

\begin{figure*}
\includegraphics{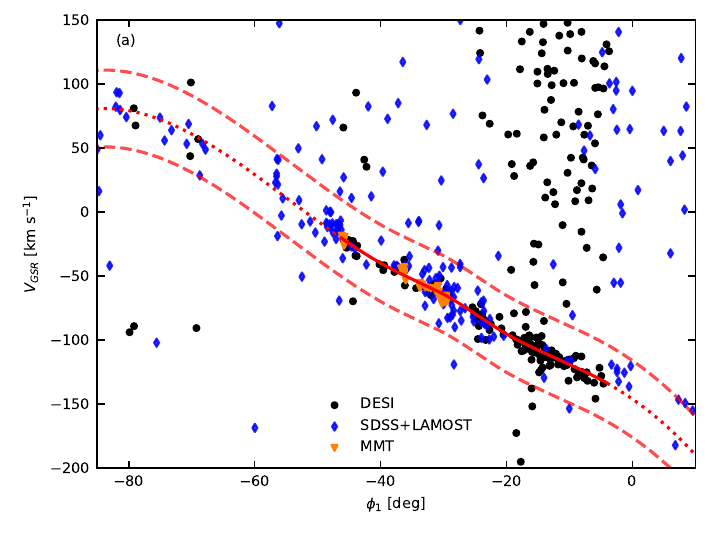}
\includegraphics{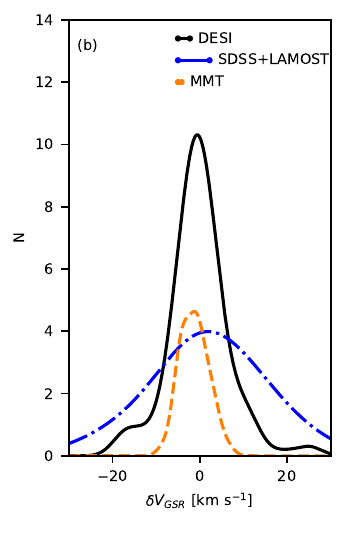}
\caption{(a) Line-of-sight velocities (corrected for solar reflex motion and transformed to GSR frame) for stars from DESI  (black points), SDSS \citep{sdssDR9} or LAMOST \citep{LAMOST} ~(blue diamonds), MMT-Hectochelle \citep{Bonaca_2020} (orange triangles). All stars were selected with the parallax, proper motion, and CMD cuts discussed in the text and include only stars with $|\delta\phi_2|<3^\circ$. The stream is clearly visible  {between the red dashed curves. The solid red curve shows the spline fit obtained using the {\tt Stan} package (see text), while the dotted curves show approximate (``by-eye'') splines in the regions of the stream without significant DESI measurements}. All DESI radial velocities and [Fe/H] measurements in this figure are given in Table~4 of Appendix~\ref{sec:datatable}. (b) Kernel density distributions of stars with velocities within $\pm 30$\kms\ (i.e. between dashed red lines) from the red stream track in left panel ($-85^\circ< \phi_1<20^\circ$). The kernel band widths are  2.74\kms\ for DESI (black curve), 7.25\kms\ for SDSS+LAMOST (blue dot-dashed curve) and 1.5\kms\ for MMT (orange dashed curve).  {For the DESI and SDSS+LAMOST data, the bandwidths correspond to the median radial velocity uncertainty. For the MMT dataset the uncertainty is 0.7\kms.}
}
\label{fig:rv_phi1_all}
\end{figure*}

With the stream  defined by the cuts discussed in the previous subsections, we now examine the radial velocities of the selected stars. Figure~\ref{fig:rv_phi1_all}~(a)  shows the solar reflex-motion corrected line-of-sight velocities in the Galactic Standard of Rest frame $V_{GSR}$ as a function of $\phi_1$ for stars in the GD-1 region selected by parallax, proper-motion (as shown in Figure~\ref{fig:mu1mu2}),  CMD cuts discussed in the text and include only stars with $|\delta\phi_2|\leq 3^\circ$.  {This value of $|\delta\phi_2|$ is motivated by the fact that in EDR, DESI only observed within $3^\circ$ from the stream track. Future data releases will include much wider coverage.} Stars observed by DESI from the EDR catalog \citep{desi_edr} are shown as black points. Stars observed by SDSS \citep{sdssDR9} (blue)  DR9\footnote{\url{https://www.sdss3.org/dr9/}} and   LAMOST DR8 \footnote{\url{http://www.lamost.org/dr8/}} are shown by blue symbols. The 43 stars identified as GD-1 members using MMT-Hectochelle spectra \citep{Bonaca_2020} are shown by orange symbols. The selection of \citep{Bonaca_2020} only considered stars with velocities within 7\kms\ from the stream track or the spur as GD-1 members, but we do not expand on this selection since few of their stars belong to the cocoon region we discuss later.

All DESI-EDR measurement in Figure~\ref{fig:rv_phi1_all}~(a) are given in a machine readable FITS file available at \url{https://zenodo.org/records/11638330}, the first 3 rows of the table are shown in Table~4 in Appendix~\ref{sec:datatable}.

We note that 18 stars from the SDSS and LAMOST datasets were also observed by DESI. Where there were duplicates we selected the measurement with the smaller velocity uncertainty. Except for one star DESI velocity uncertainties were always smaller than uncertainties from SDSS and LAMOST. There were no overlaps with GD-1 members in the MMT sample of \citet{Bonaca_2020} primarily because the DESI-EDR observations did not have much coverage of the spur area. Of the remaining 153 SDSS+LAMOST stars 14 did not have [Fe/H] measurements and were therefore excluded.

In summary DESI has added 115 new spectroscopically confirmed radial velocities and improved RV measurements for 17 previously observed members.

A spline fit to the stream track is shown  {with a red curve (partially solid, partially dotted)} (see below for details of how this fit is obtained) along with dashed lines showing $\delta V_{GSR} = \pm 30$\kms\ from the stream track. Figure~\ref{fig:rv_phi1_all}~(b) shows kernel density estimates of $\delta V_{GSR}$  (from the stream track in the left panel) of stars observed by DESI (black), SDSS+LAMOST (blue) and MMT (orange). All stars within $-85^\circ< \phi_1<20^\circ$ are shown.  A clear peak is visible identifying the GD-1 stream. 
The  broader histogram corresponding to stars from SDSS+LAMOST reflects their larger measurement uncertainties, while the narrower MMT distribution reflects both the small uncertainties and the more restrictive selection of $|\delta V_{GSR}|\leq7$\kms\ imposed by \citet{Bonaca_2020}.

 {In order to obtain the stream  track  we use a mixture modeling approach that properly accounts for the uncertainties in the radial velocity measurements and background contamination.} We use the approach previously adopted for other streams in \citet{koposov2019_orphan, koposov_2023_OC}. 
Focusing on the region along the stream of $-55^\circ \leq \phi_1 \leq 0^\circ$, where the density of the stream and number of DESI RV measurements is the highest, we construct a model for the radial velocity distribution as a function of $\phi_1$ (only DESI measurements are used in determining the spline).  We model the distribution as a Gaussian mixture where the center of the Gaussian representing the GD-1 stream is allowed to vary with $\phi_1$. The other Gaussian represents the background contamination. The variation of the radial velocity of stream stars is represented by a spline with 6 knots. The velocity dispersion of the stream is also allowed to vary and is represented by a spline with 4 knots. The mean and the velocity dispersion of the background contamination are fitted parameters. The specific likelihood function for the radial velocity  {$v$ conditional on $\phi_1$ is given by,
\begin{equation}
\begin{aligned}
P(v|\phi_1) = & (1-f(\phi_1))\mathcal{N}(v|v_{bg} + d_{bg}\phi_1, \mathcal{S}_{bg}) + \\
 & f(\phi_1)\mathcal{N}(v|\mathcal{V}(\phi_1), \mathcal{S}_{\mathcal{V}}(\phi_1), \label{eq:stan_eq2} 
 \end{aligned}
\end{equation}
which is identical to  equation 2 of \citet{koposov_2023_OC}. In the above equation
$f(\phi_1)$ is the mixing fraction, $\mathcal{V}(\phi_1)$ is the velocity of the stream, and $\mathcal{S}_{\mathcal{V}}(\phi_1)$ is the velocity dispersion of the stream, all of which are represented as cubic splines with the values at
the knots being model parameters. $v_{bg}$ is the mean velocity of the Gaussian describing the background, $d_{bg}$ is the gradient of the mean velocity of the background with respect to $\phi_1$, and $\mathcal{S}_{bg}$ is the velocity dispersion of the background. }

The model is implemented using the {\tt Stan} probabilistic programming language \citep{Carpenter_etal_2017} and is sampled using the {\tt CMDstanpy} package. The splines were implemented using the {\tt stan-splines} package using code very similar to that provided in the supplementary material in  \citet{koposov_2023_OC}. When computing the likelihoods of individual stars, we account for their radial velocity uncertainties by adding them in quadrature to the intrinsic velocity dispersions. The priors on model parameters are mostly non-informative. The velocity prior for individual spline knots is a normal distribution $\mathcal{N}(0,300)$; the  prior on the logarithm of the stream velocity dispersions at knots is  $\mathcal{N}(1.6, 1.6)$ which is also uninformative. For the background contamination the priors on the mean velocity, velocity dispersion and velocity gradient are also normal distributions with large standard deviations.
 {We note that although the method yields a measurement of the velocity dispersion along the stream, the main purpose of this analysis  is to robustly obtain the stream track in the presence of the background.}

\begin{figure}[t]
\centering
\includegraphics[trim = 0 10 0 0, clip]{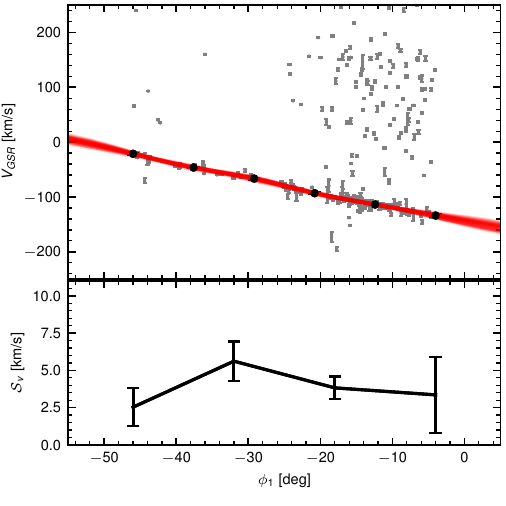}
\caption{Top: DESI-observations in the GD-1 region. Middle measurement of velocities at spline knots is shown by black points with error bars. The red curves show the samples from the posterior distribution. The grey points show the stars used to obtain the spline fit. Bottom: The velocity dispersion of the GD-1 stream as measured by the spline model. The error bars show the $1\sigma$ uncertainties  in the velocity dispersion measurements derived from the posterior samples at four points along the stream.}
\label{fig:rv_disp_spline}
\end{figure}

Figure~\ref{fig:rv_disp_spline}~(top) shows  the measured radial velocity $V_{GSR}$ vs. $\phi_1$ for all DESI targets within $\pm 3^\circ$ from the stream track (grey points). The uncertainties on the velocities are not visible as they are $\lesssim $ 15 km/s. Red lines show posterior samples from radial velocity curves.  {Spline knots (obtained with Stan) are shown as black dots.}

The lower panel of Figure~\ref{fig:rv_disp_spline} shows the velocity dispersion along the stream  in the region $-46^\circ \leq \phi_1 \leq -4^\circ$. The error-bars show the   {1$\sigma$ uncertainties defined as the 16th-84th percentile credible intervals} determined from the posterior samples  {(also given in Table~2)}.
The velocity dispersion   {everywhere  is  between 2.36 and 5.47\kms\ with the uncertainties of $\pm 0.7-\pm2.5$\kms.}

 We will return to the issue of the radial velocity dispersion in Section~\ref{sec:cocoon}. 

\begin{table}

\centering
\begin{tabular}{ccc}
\multicolumn{3}{c}{{\bf Table 2: Stream velocity dispersions }}\\
\hline 
$\phi_1$ &  $\sigma_{V_{GSR}}$ & 1$\sigma$ uncertainty \\
 degrees  &\kms             & \kms      \\
\hline
-46.0   & 2.36  & 1.24 \\
-32.0   & 5.47  & 1.26 \\
-18.0   & 3.78  & 0.73 \\
-4.0    & 2.72  & 2.45 \\
 \hline \\
\end{tabular}
\label{tab:stan_disp}
\end{table}

The {\tt Stan} spline fit above is obtained by using only DESI velocity measurements (since these have the smallest uncertainties)  {and is shown by the solid red curve in   Figure~\ref{fig:rv_phi1_all}. Beyond the range of $\phi_1$ values modeled by the {\tt Stan} spline, we  select knots by eye, rather than by using the median/mean velocity in bins primarily because the there are few DESI detections in these regions and the outliers are not symmetrically distributed relative to the stream track  (see Fig.~\ref{fig:rv_disp_spline}). The 3-year DESI sample has substantially more observations available and will allow for a more robust analysis along the entire length of the stream.}
 
\subsection{Metallicity distribution of stream stars \label{sec:metallicity}}

\begin{figure}[t]
\centering
\includegraphics[trim = 0 0 0 10,clip]{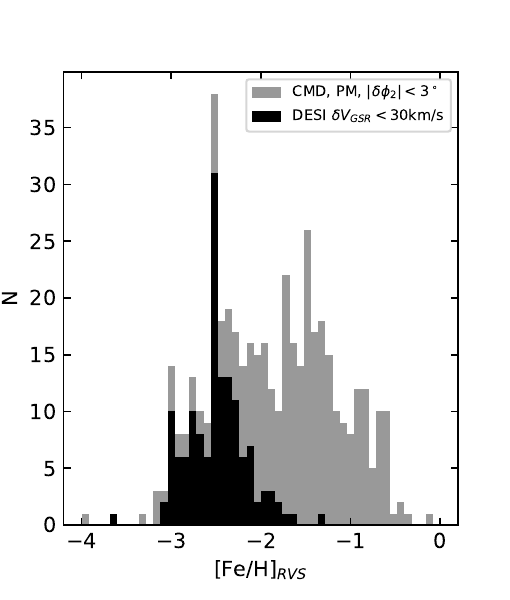}
\caption{[Fe/H] distribution for stars in the GD-1 stream  using the  proper motion cuts, parallax cut, CMD cut in the GD-1-region in previous figures (grey). Black histogram shows stars with the additional constraint that they have $\delta V_{GSR}< 30$\kms\, from the stream track in Figure~\ref{fig:rv_phi1_all}(a). Only DESI measurements are used in this figure.}
\label{fig:feh}
\end{figure}

Figure~\ref{fig:feh} shows the distribution of DESI [Fe/H] measurements (from the RVS pipeline) for all stars (grey) selected using the proper motion, parallax, and CMD cuts  using the selections described in Sections~\ref{sec:pm} and \ref{sec:CMD}. Stars with the additional constraint that they lie within 30~\kms\ from the stream track in Figure~\ref{fig:rv_phi1_all} (Section~\ref{sec:radial_vel}) are shown in black. It is clear from the black histogram in Figure~\ref{fig:feh} that only one star has [Fe/H] $> -1.5$ and one has [Fe/H]$< -3.5$.  We exclude these stars bringing the sample from DESI to { 151 (115 new, 36 improved)}. With these metallicity cuts 186 GD-1 stars from SDSS+LAMOST remain, along with 43 stars from MMT \citep{Bonaca_2020}. 

 {As discussed in Section ~\ref{sec:data}  comparisons with 507 DESI stars in also observed by APOGEE indicate that the RVS metallicity has a median offset [Fe/H]$_\mathrm{DESI}-$[Fe/H]$_\mathrm{APOGEE}=0.06^{+0.13}_{-0.18}$~dex. However, despite this small offset, we note that \citet{Cooper2023overview} showed that [Fe/H] measurements obtained with DESI for several GCs (M13, M92, NGC~5053) showed significant spreads of 0.3 dex or greater \footnote{compared to typical spreads of $0.03-0.05$~dex for GCs \citep{Li_2022_S5}.}. Furthermore they often had mean metallicities that were systematically lower by 0.2 to 0.4 dex relative to literature values.
Given these rather large observed  spread of 0.2-0.3dex in the [Fe/H] values recovered by the RVS pipeline we caution readers against infering too much from the spread in [Fe/H] for GD-1 members reported in this paper. 

Despite these concerns with current DESI metallicity measurements the mean [Fe/H]$=-2.49\pm0.25$ for all stars in the black histogram is } consistent with values from previous works: [Fe/H] = -2.24$\pm 0.21$ \citep{Malhan_2019_shape}, [Fe/H]=-2.3$\pm 0.1$ \citep{Bonaca_2020},  {and [Fe/H] = -2.49$\pm 0.03$ } \citet{Martin_etal_2022Pristine} and [Fe/H] = -2.47 \citep{Ibata_2023}.

\begin{figure*}[thb]
\begin{centering}
\includegraphics[width=\textwidth]{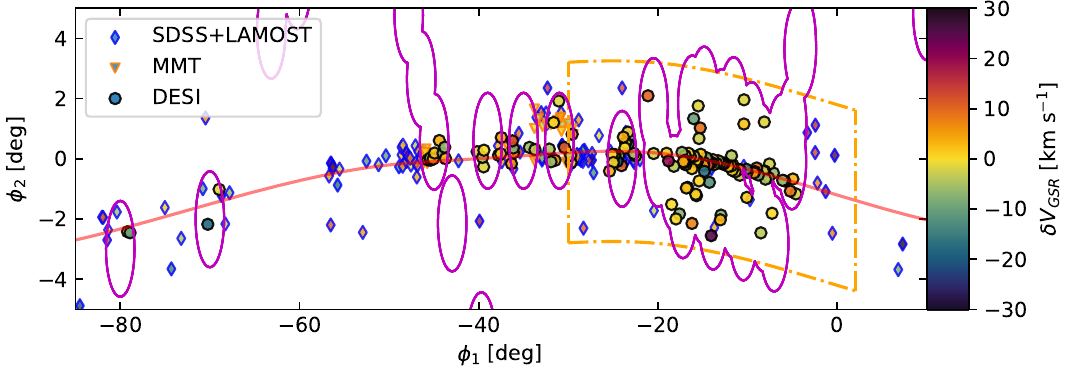}
\caption{
Stars in the GD-1 region with $|\delta \phi_2|<3^\circ$ from the stream track with parallax, CMD, proper motion and metallicity  ($-3.5<$[Fe/H]$<-1.5$)) cuts, colored by $\delta V_{GSR}$. Points show stars observed by  DESI-EDR (circles),  SDSS-DR9+LAMOST-DR8 (diamonds),  and MMT-Hectochelle (triangles). The orange box shows the region we analyze for the cocoon (where most of the DESI observations lie). Boundaries of DESI tiles (fields) are shown by magenta curves.}
\end{centering}
\label{fig:gd1sel}
\end{figure*}

\begin{figure}[thb]
\begin{centering}
\includegraphics{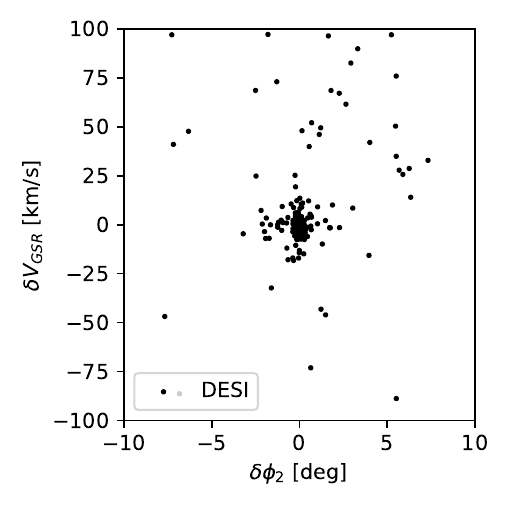}
\caption{Distribution of DESI stars in  $\delta V_{GSR}$ and $\delta\phi_2$. 
The centrally concentrated 
component of the GD-1 stream and the broader cocoon 
component are clearly visible.}
\end{centering}
\label{fig:cocoon_largefov}
\end{figure}

\section{Identification and characterization of the cocoon \label{sec:cocoon}} 

As discussed in Section~\ref{sec:intro}, an extended distribution called a ``cocoon'' has been identified surrounding the GD-1 stream and a few other tidal streams. While the origin of such cocoons is still uncertain (see Section~\ref{sec:discussion}), with the increased number of radial velocities provided by DESI it is now possible to identify and characterize the GD-1 cocoon in greater detail.

Our final selection (after imposing the parallax, proper-motion, CMD, RV and metallicity cuts previously described and with $|\delta\phi_2|\leq3^\circ$),  can be seen in Figure~\ref{fig:gd1sel}. This plot only shows observations with full 6-D phase space and metallicity information.   \citet{Malhan_2019_cocoon} identified the GD-1 cocoon primarily in the region $-10^\circ < \phi_1 <0^\circ$ based on Gaia positional and proper motion data.  DESI spectroscopic measurements confidently confirm the existence of the cocoon as a broad component in $\phi_2$ extending between $-30^\circ< \phi_1  < 2^\circ$. 

We now analyze the phase space distribution of the stars in the region bounded by the orange box. In this work we restrict our analysis to this range of $\phi_1$ since it is where the majority of the DESI spectroscopic measurements exist. We  do not claim that this is the full extent of the cocoon.  Spectroscopic observations at other positions along $\phi_1$ and extending to larger distances perpendicular to the stream are required to confidently assess the full extent of the cocoon and will be presented in future work. The cocoon might well extend further down to $\phi_1 \sim -38^\circ$ but we have few DESI-EDR observations in this region. This region also includes the `spur' feature (seen as orange triangles above the main stream track) and  we choose to ignore it here since a specific model has been postulated for the origin of this structure \citep{Price-Whelan_2018,Bonaca_2019}.

Figure~\ref{fig:cocoon_largefov} 
shows a broader velocity range of stars in the orange box in Figure~\ref{fig:gd1sel} to illustrate that the velocity range that we use ($\pm 30$\kms) does indeed completely capture the cocoon region around the narrower GD-1 stream. 
Since Figure~\ref{fig:cocoon_largefov} only includes data observed by DESI, the extent of the cocoon in $\phi_2$  in this figure may be underestimated.

\begin{figure}
\begin{centering}  \includegraphics{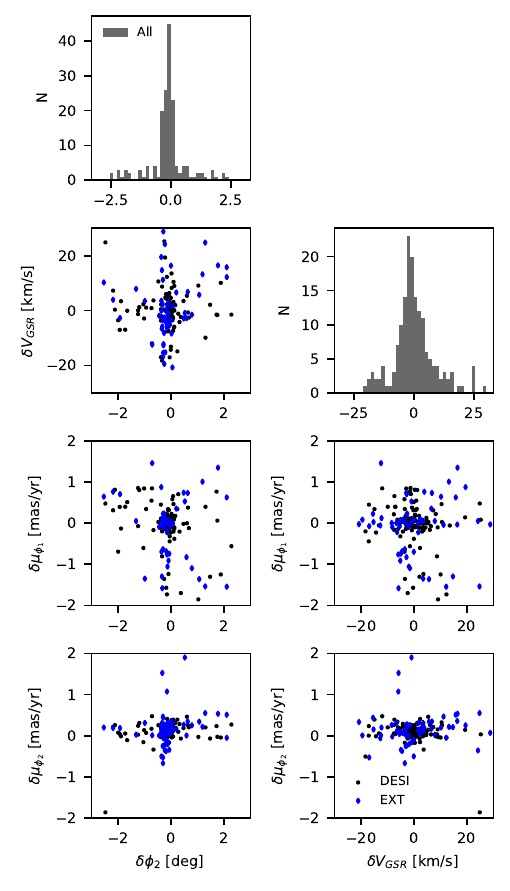}
    \caption{Stars from region in the orange box in Fig.~\ref{fig:gd1sel} with the same selections. DESI observations are in black, SDSS+LAMOST+MMT in blue. In each scatter plot and histogram it is clear that there are both narrow and broad components in $\delta \phi_2$, $\delta V_{GSR}$ and $\delta\mu_{\phi_1}$. The overall width in $\delta\mu_{\phi_2}$ (perpendicular to the stream track) is smaller than in  $\delta\mu_{\phi_1}$ (along the stream track).}
\end{centering}
\label{fig:cocoon_various}
\end{figure}

Figure~\ref{fig:cocoon_various} shows various phase space projections of stars in the orange box in Figure~\ref{fig:gd1sel} relative to the stream track (as identified by the  4 stream-track splines previously defined in Table~1). In Figure~\ref{fig:cocoon_various} the left column shows (top to bottom): a histogram of $\delta\phi_2$  and scatter plots of $\delta V_{GSR}$,  $\delta \mu_{\phi_1}$, $\delta \mu_{\phi_2}$ vs. $\delta\phi_2$. In all plots in this column a centrally peaked distribution in $\delta{\phi_2}$ is clearly distinguishable from a more extended distribution in $\delta \phi_2$ corresponding to the cocoon. The plots in the right hand column show (top to bottom): a histogram of  $\delta V_{GSR}$ and scatter plots of  $\delta\mu_{\phi_1}$  and $\delta\mu_{\phi_2}$ vs. $\delta V_{GSR}$. We also note that the plots  show  evidence for a greater spread in $\delta V_{GSR}$ and $\delta\mu_{\phi_1}$ (two components of motion along the stream) than in $\delta\mu_{\phi_2}$ (motion perpendicular to the stream direction).  {In particular we note that the significant  spread in $V_{GSR}$ at small $|\delta\mu_{\phi_1}|$, likely arises from cocoon stars in front of and behind the thin stream, evidence that the cocoon is probably a 3D cylindrical structure around the thin stream.}

\subsection{Multi-Gaussian Analysis \label{sec:univariate}}
We wish to separate the ``thin stream'' and ``cocoon'' components in order to individually examine their phase space distributions.  We use a  Gaussian mixture model  to fit observed quantities  with two  univariate Gaussians (representing the displacement of stars from the stream track in  $\delta\phi_2$). One Gaussian represents the  ``thin stream'' component 
$\mathcal{N}_1(\bar{x}_1,\Sigma_1)$ while the other represents the cocoon  $\mathcal{N}_2(\bar{x}_2,\Sigma_2)$, with an unknown mixing fraction $Q$.
We adapt the formalism of Equation~17 of \citet{Hogg_etal_2010} to write the likelihood function as follows, where $x_i$ refer to individual values of   $\delta\phi_2$. We assume that positional uncertainty on all $\delta_{\phi_2}$ values is zero since  the Gaia positional uncertainty at $G=21$ is  0.1~mas which is much smaller (by a factor $10^7$) than the spread in $\delta\phi_2$.
$Q$ is the fraction of stars in Gaussian component $\mathcal{N}_1$ (``thin stream'').  Assuming each component is well described by a Gaussian, the likelihood function can be written as,
\begin{equation}
\begin{aligned}
\mathcal{L}  =  & \prod_{i=1}^{N} Q\mathcal{N}_1(x_i|\bar{x}_1,\Sigma_1) + (1-Q)\mathcal{N}_2 (x_i|\bar{x}_2,\Sigma_2),  \\
\mathcal{L}   = & \prod_{i=1}^{N}\frac{Q}{\sqrt{2\pi\Sigma_1^2}}\exp{\left(-\frac{(x_i-\bar{x}_1)^2}{2\Sigma_1^2}\right)} \\
           &     + \frac{(1-Q)}{\sqrt{2\pi\Sigma_2^2}}\exp{\left(-\frac{(x_i-\bar{x}_2)^2}{\Sigma_2^2}\right)}.
           \label{eq:likelihood1}
\end{aligned}
\end{equation}

We estimate the five parameters in the above model ($\bar{x}_1$, $\Sigma_1$, $\bar{x}_2$, $\Sigma_2$, $Q$) in two different ways.
First, we use direct maximization of the likelihood in Equation~\ref{eq:likelihood1} with {\tt scipy.optimize} package. We determine the distribution of the best-fit parameters by bootstrapping the observed sample 1000 times.

Second, we generate the posterior distribution of the model parameters using the ensemble Markov Chain Monte Carlo sampling method {\tt emcee} \citep{emcee}. We adopt uniform flat priors for all the parameters ($\bar{x}_1$, $\Sigma_1^2$, $\bar{x}_2$, $\Sigma_2^2$, $Q$). The posterior samples are generated with 64 walkers until the convergence of the mean parameter values to 1\%. Since the maximum autocorrelation length of the chain $N_{\rm corr} \approx 80$ steps, we discard the first 1600 samples as ``burn-in'' and thin the remaining chain by a factor $N_{\rm corr}$. It results in
about 6000 independent chain steps.

Both direct maximization of the likelihood and the MCMC method give nearly identical result within the uncertainties. Both methods also yield similar means and standard deviations for the two Gaussians in $\delta\phi_2$ and hence we report the results of the Bayesian MCMC analysis. The resulting mean ($\overline{\delta\phi_2}$) and standard deviation ($\Sigma_{\delta\phi_2}$) of the Gaussians are reported in Table~3 (top 3 rows, of 2nd \& 3rd columns). Uncertainties on all parameters  in this table are $1\sigma$ uncertainties determined from the 16th and 84th percentile credible intervals of the posterior probability distributions.

\begin{table*}
\centering
\begin{tabular}{lrrcc}
\multicolumn{5}{c}{{\bf Table 3: Results of Univariate and Bivariate multi-Gaussian analysis }}\\
\hline
 {\bf Parameter} & \multicolumn{2}{c}{\bf Univariate }&  \multicolumn{2}{c}{\bf Bivariate}\\
    \phantom{xxxxx}      & \multicolumn{1}{c}{Thin}  & \multicolumn{1}{c}{Cocoon} & \multicolumn{1}{c}{Thin}  & \multicolumn{1}{c}{Cocoon} \\
\hline 
$\overline{\delta\phi_2}$ [deg]    & $-0.118\pm 0.017$  & $-0.149\pm 0.161$ & $-0.111\pm0.017$ & $-0.161\pm0.157$ \\ 
$\Sigma_{\delta\phi_2}$ [deg]      & $0.149 \pm 0.014$  & $1.265 \pm 0.122$ & $0.149\pm0.014$ & $1.252\pm0.121$  \\
$\overline{\delta V_{GSR}}$ [\kms] & \ldots             & \ldots            &  $-1.19\pm0.52$ & $1.67\pm1.21$    \\
$\Sigma_{\delta V_{GSR}} [\kms]$   & \ldots             & \ldots            & $3.090\pm0.76$   & $7.81\pm1.01$     \\
$Q_{\rm thin}$ & $0.63\pm0.05$     & \ldots             &  $0.61\pm0.05$            & \ldots \\
\hline

\end{tabular}
\label{tab:multi-gaussian_results}
\end{table*}
 
\begin{figure}[t!]
\begin{centering}
    \includegraphics[trim = 0 10 0 5, clip]{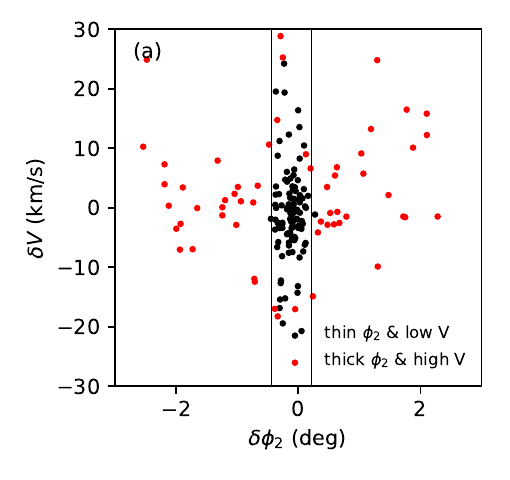}
     \includegraphics[trim = 0 10 0 5, clip]{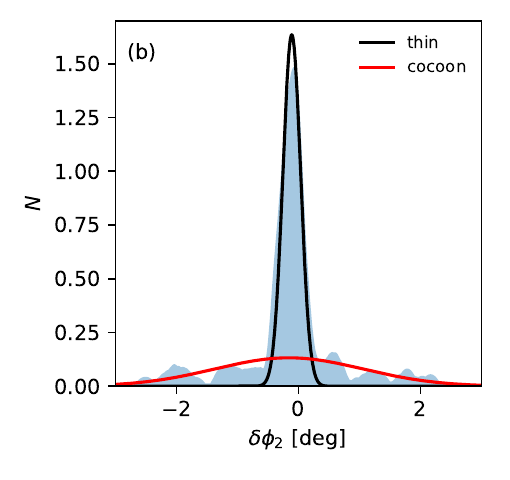}
\caption{(a) Relative radial velocity $\delta V_{GSR}$ vs. $\delta\phi_2$.  Points  show the thin and lower velocity dispersion stream stars (black) and thick and higher velocity dispersion cocoon stars (red) as determined by the bivariate double-Gaussian model. The vertical lines are at $\overline{\delta\phi}_{2,thin}\pm \Sigma_{\delta\phi_{2,thin}}$ and include the majority of the thin stream stars identified as members of the thin component by both position and velocity. (b) Kernel density estimates of in $\delta\phi_2$  for all stars in the orange box in Figure~6 with Gaussians representing the  thin stream (black curve) and cocoon (red curve) as estimated by the bivariate analysis.}
\end{centering}
\label{fig:cocoon_classify}
\end{figure}

\begin{figure*}[t!]
\begin{centering}
    \includegraphics[trim =0 50 0 0, clip, width=0.70\textwidth]{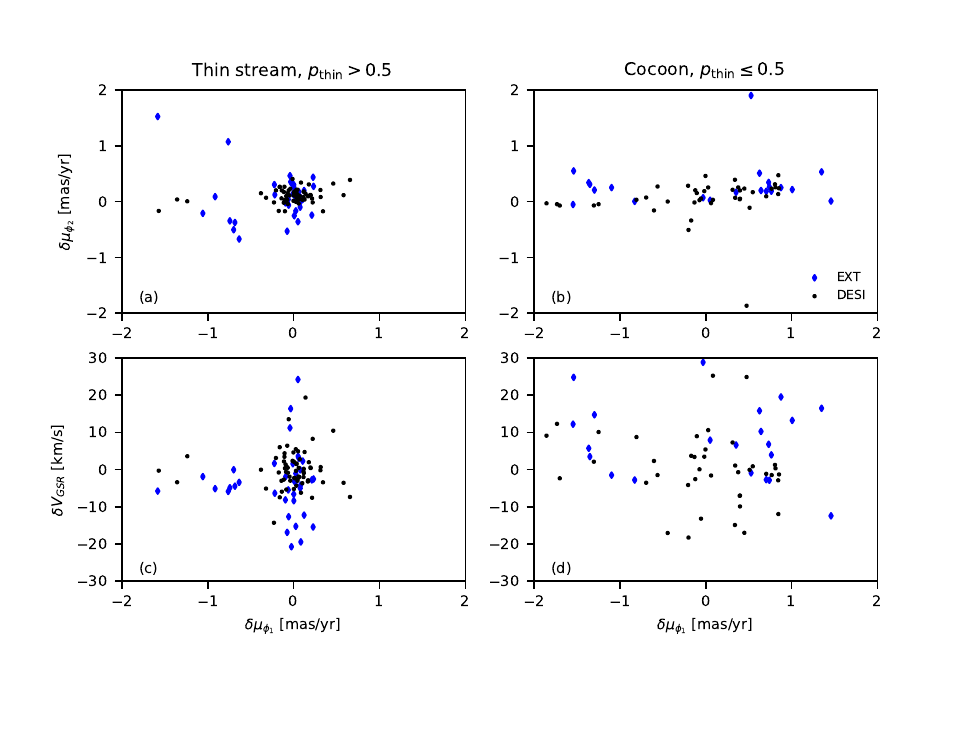}
\caption{Relative radial velocity and both components of proper motions of the stars from stream (left column) and cocoon (right column) using membership probabilities from the bivariate Gaussian model to define thin and cocoon components.}
\end{centering}
\label{fig:stream_cocoon_velspace}
\end{figure*}

Next we jointly fit the $\phi_2$ and $V_{GSR}$ distributions  with two bivariate-Gaussians described by the likelihood function,
\begin{eqnarray}
\mathcal{L}  & =  & \prod_{i=1}^{N} Q\mathcal{N}_1(x_i,y_i, |\bar{x}_{1}, \bar{y}_{1}, \Sigma_{x1},\Sigma_{y1}, \sigma_{xi}, \sigma_{yi}) + \nonumber \\
 & & (1-Q)\mathcal{N}_2(x_i,y_i, |\bar{x}_{2}, \bar{y}_{2}, \Sigma_{x2},\Sigma_{y2}, \sigma_{xi}, \sigma_{yi}),
 \label{eq:likelihood2}
\end{eqnarray}
where both $\mathcal{N}_1$ (thin stream) and  $\mathcal{N}_2$ (cocoon) are  represented by 2D (bivariate) Gaussians  $\mathcal{N}(\delta\phi_2, \delta V_{GSR})$.  

The results of the Bayesian MCMC analysis of the bivariate Gaussian model (9 parameters and their uncertainties) are also given in Table~3.  

Both the univariate and bivariate analyses give nearly identical widths in $\delta\phi_2$  for the `thin stream' and cocoon. The standard deviation of the Gaussian fit to the thin component  
$\Sigma_{\delta\phi_2,thin} = (0.149\pm0.014)^\circ$ implies a FWHM of the ``thin stream'' component in $\phi_2$ to be 2.355$\times \sigma_{\phi_2, thin} =0.351^\circ\approx 55$~pc (at a  mean distance of 9~kpc) while the cocoon component is much broader with standard deviation
$\Sigma_{\delta\phi_2,\mathrm{cocoon}} = (1.252\pm0.121)^\circ$ or FWHM $\approx  2.948^\circ \approx 463$~pc.
For the distribution in $V_{GSR}$ the bivariate analysis gives velocity dispersions
$\Sigma_{V_{GSR},\mathrm{thin}}=(3.090\pm 0.76)$~\kms\ and $\Sigma_{V_{GSR},\mathrm{cocoon}}=(7.81\pm 1.01)$~\kms\ 
for the thin stream and cocoon components respectively, with the fraction of stars in the thin stream $Q = 0.61\pm0.05$. 

The standard deviations in $\delta\phi_2$   of the thin stream and cocoon obtained both by
only modeling $\delta\phi_2$ versus by jointly modeling $\delta\phi_2$ and $\delta V_{GSR}$ yields nearly identical results for the angular widths of the thin stream and cocoon. Both methods also give $Q \sim 0.62$ for the fraction of stars in the `thin stream' component.

Figure~\ref{fig:cocoon_classify}~(upper panel)  shows the stars belonging to the ``thin stream''  with probability $p_{thin}\geq0.5$ as determined by the bivariate analysis considering  both  position and velocity $\delta V_{GSR}$. The vertical lines are located at $\overline{\delta\phi}_{2,thin}\pm \Sigma_{\delta\phi_{2,thin}}$ from the bivariate analysis in Table~3 and include the majority of points determined to be members of the ``thin'' component by both univariate and bivariate analyses. 
The vertical lines are not used in subsequent analysis and only meant to guide the eye. Figure~\ref{fig:cocoon_classify}~(lower panel) shows the kernel density estimate of all the stars in the orange box in Figure~\ref{fig:gd1sel}.
Overplotted as black and red curves are the Gaussians in $\delta\phi_2$ for the thin stream and cocoon respectively (as determined by the bivariate Gaussian analysis). 

Figure~\ref{fig:stream_cocoon_velspace}  shows various phase-space projections of the thin stream (left) and the broader cocoon component (right) with membership defined by the individual probabilities obtained from the bivariate analysis.  It is clear that the cocoon has a much greater spread in $\delta\mu_{\phi_1}$ and $\delta V_{GSR}$ than in $\delta\mu_{\phi_2}$. Although the ``thin stream'' component defined in this way has a very narrow central component in $\delta\mu_{\phi_1}$, some stars show significant velocity and proper motion deviations. The ``thin stream'' component also appears to have both narrow and broad components in $\delta V_{GSR}$. Both the thin stream and cocoon components have narrow distributions in $\delta\mu_{\phi_2}$, but the cocoon has broad distributions in both $\delta V_{GSR}, \delta\mu_{\phi_1}$.  {As stated previously the spread in $\delta V_{GSR}$ could be due to high velocity cocoon stars in front of and behind the thin component of the stream. In addition, over this range of $\phi_1$, the distance gradient implies that the stream is at an angle of $\sim 30^\circ$ to our line-of-sight.  Thus the increased dispersion in  $\delta V_{GSR}$ and $\delta\mu_{\phi_1}$ but not in $\delta\mu_{\phi_2}$ suggests that there has been greater heating {\it along} the stream than perpendicular to the stream.}

\subsection{Accounting for contamination by background stars \label{sec:background}}

In the above analysis we have not explicitly modeled the background population. Adding a third bivariate Gaussian component to represent contamination by a background population adds 5 additional parameters. Given the small number of stars in total in this region of the stream we did not attempt such a model. Instead we use the fact that both of the previous analyses yielded consistent values for the width of the thin stream. We use the probabilities of membership to the `thin stream' and cocoon based on the univariate analysis in Section~\ref{sec:univariate} to define each component. We then model each component's velocity distribution with a univariate Gaussian  {for that} component and represent the background by a much broader Gaussian in velocity truncated at $\delta V_{GSR} = \pm 30$\kms\ (the velocity cut shown by the dashed lines in Fig.~\ref{fig:rv_phi1_all}a). The mean  { velocity and dispersion of each component (thin stream or cocoon) and} the background contamination are fitted parameters. The likelihood function is similar to that given in Equation~\ref{eq:likelihood1}.  {Therefore we repeat the exercise of modeling a main component and a broad  truncated Gaussian contaminant {\bf twice}, once for the ``thin stream'' members and again for the ``cocoon'' members.}

As one would expect if some fraction of the higher velocity stars in the `thin stream' or cocoon are classified as belonging to the background, the velocity dispersions of these two components would decrease. 

 {When this analysis is applied to the members identified with the physically ``thin'' component (stars colored black in Figure~\ref{fig:cocoon_classify}} the model yields a  {thin stream} velocity dispersion of  {$2.85\pm0.6$} \kms\ and the fraction of stars assigned to the background was $0.12\pm 0.05$. This velocity dispersion for the thin stream component is consistent (within the uncertainties) to the dispersion obtained from the bivariate multi-Gaussian analysis in the previous section. 

When applied to the members of the  {physically broad} cocoon component  {(stars colored red in Figure~\ref{fig:cocoon_classify}} we find that the cocoon has a smaller velocity dispersion of  {$4.84\pm1.13$} \kms\ and the background fraction in the cocoon is also estimated to be 12\%. 

To assess whether this background fraction is reasonable we counted the number of stars (after applying the same cuts used to select stars in the orange box in Fig.~\ref{fig:gd1sel}) in three separate off-stream regions with the same size as the orange box. The number of stars in these 3 regions compared to the number of cocoon stars varied from 4-8\% suggesting that the number of background stars in the cocoon could be somewhat lower than in the above estimate of 12\%. 

The  analyses in this subsection and the previous one both imply that the physically thin stream has a  velocity dispersion of $\sim 2.85-3.09$\kms, while the cocoon has a velocity dispersion between 4.84-7.81\kms. Additional data from future DESI data releases will  enable us to determine if the cocoon is an extended structure found along the entire stream or if it is more localized. It will also enable us to derive a more robust estimate of the velocity dispersion of the cocoon.

 {We remind the reader that the {\tt Stan} analysis in section~\ref{sec:radial_vel}, which was primarily intended to derive the stream track in $V_{GSR}-\phi_1$ space while properly accounting for background contamination, also resulted in velocity dispersions in the $-32\leq \phi_1\leq -4$ of $5.47\pm1.26$ to $2.72\pm2.45$\kms, which is consistent with the above estimates (see Table~2).}

Our measurement of the dispersion of the `thin'  stream component ($\sigma_{V_{GSR},\mathrm{thin}} \sim 3$~\kms) is consistent with previous estimates of both other GC streams and the GD-1 stream itself \citep{Gialluca_2021}. In contrast, the larger velocity dispersion of the broader cocoon component of $\sigma_{V_{GSR},\mathrm{cocoon}} \sim 5-8$~\kms\ is comparable to the velocity dispersion of classical dwarf spheroidal galaxies like Carina, Draco, Fornax, Leo I etc. \citep{Simon_2019}. This measurement is also comparable to the velocity dispersion of  other dwarf galaxy streams \citep[e.g.,][]{Li_2022_S5} which have velocity dispersions that range from 4-20\kms\ with a mean value of around 11\kms\ and are significantly larger than the dispersions of known  GC streams  \citep[e.g.,][]{Li_2022_S5}. 

\subsection{Metallicity of thin stream and cocoon components \label{sec:coc_metallicity}}
Another way to assess if the cocoon velocity dispersion in Section~\ref{sec:univariate} is inflated by background/foreground halo stars is to examine the metallicity distributions of stars in the two kinematically and spatially identified components. 
\begin{figure}
\begin{centering}
    \includegraphics[trim=10 15 10 10, clip]{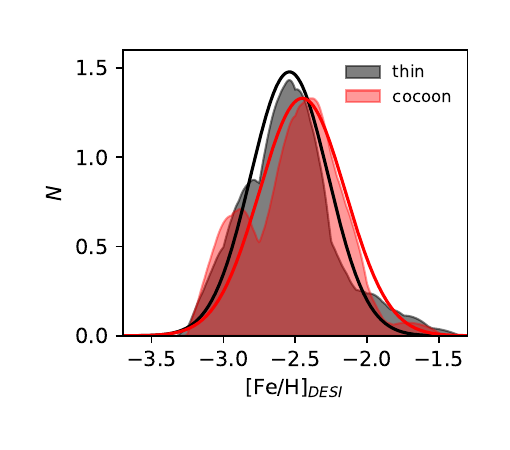}
    \caption{Kernel density estimates (with an Epanechnikov kernel) of [Fe/H] as reported by the DESI-RVS pipeline for the `thin stream' component  (black/grey) and the cocoon component (red).  The average random error on the [Fe/H] measurements is $\sim 0.2$ dex and is also the band-width of the kernel density estimate. Overplotted are the best-fit Gaussians (see text for details).}
    \label{fig:cocoon_metallicity}
    \end{centering}
\end{figure}

Figure~\ref{fig:cocoon_metallicity}  shows  the  [Fe/H] distribution functions for the thin stream component (grey/black) and the cocoon component (red), where the membership probability from the bivariate analysis was used to assign memberships. It is clear that the distribution of [Fe/H] for cocoon stars  overlaps significantly with the distribution for thin stream stars   {and the metallicity of the cocoon stars is indistinguishable from that of the thin component within the uncertainties}.

Once again assuming that the metallicity distribution for each component is described by  a Gaussian we use a Bayesian MCMC analysis (32 walkers, 1000 steps) to determine their intrinsic  means and intrinsic standard deviations while accounting for the  uncertainties in the measurements. 
The mean metallicity of the `thin stream' component is found to be $\left<\mathrm{[Fe/H]_{thin}}\right> = -2.54\pm0.04$~dex and intrinsic metallicity dispersion is $\sigma_\mathrm{[Fe/H],thin} = 0.27\pm0.03$dex.

The  cocoon component has a mean $\left<\mathrm{[Fe/H]_\mathrm{cocoon}}\right>= -2.47\pm0.06$~dex with dispersion $\sigma_\mathrm{[Fe/H],\mathrm{cocoon}} = 0.31\pm0.04$~dex. 
The means of both these distributions are  consistent with the value of [Fe/H] = $-$2.47~dex recently reported by \citet{Ibata_2023} for the entire GD1 stream. They are also significantly lower than the stellar halo metallicity of $\sim -1.5$~dex.
A Kolomogorov-Smirnov 2-sided test yields $p = 0.255$, implying that we are unable to reject the null hypothesis that the two distributions are drawn from the same parent distribution.  

A comparison with other Milky Way tidal streams \citep[see Table 1 of,][]{Li_2022_S5} shows that the inferred instrinsic metallicity dispersion  of GD-1 is significantly higher than typical GC streams which have metallicity dispersions of 0.03-0.05~dex.  This rather high inferred intrinsic metallicity dispersion of  GD-1 may be  explained by the current systematic floor of 0.3~dex in the uncertainty of [Fe/H] measurements resulting from the RV pipeline,  {as discussed in Section~\ref{sec:metallicity} and as shown in \citet{Cooper2023overview} [Fe/H] measurements in GCs obtained with DESI show significantly larger spreads of 0.3 dex or greater relative to literature values.} 
We caution that the DESI EDR  metallicity values should be treated as preliminary until ongoing improvements to the spectroscopic pipelines are completed. Some of the improvements in progress including machine learning based methods are expected to yield more accurate metallicities and abundances from the low resolution, low S/N DESI spectra  (J. Han, \& DESI Collaboration, in prep, M. Haiger,  \& DESI Collaboration, in prep). 

\section{Discussion and conclusions}
\label{sec:discussion}

\subsection{Summary of results \label{sec:summary}}
We report spectroscopic analysis of the GD-1 tidal stream using observations with Dark Energy Spectroscopic Instrument \citep{desi16a,DESI_Instr_Overview2022}.  We identify stars associated with the GD-1 stream by selecting in position $|\delta \phi_2|<3^\circ$, and proper motion $|\delta\mu_{\phi_1}| < 2$mas~yr$^{-1}$, $|\delta\mu_{\phi_2}| < 2$mas~yr$^{-1}$ relative to the stream track. Next we apply a selection using a PARSEC isochrone with [M/H] = $-$2 and age = 12~Gyr and selecting stars with $|\delta(g-r)|<0.1$ from this isochrone. After this selection we identify and fit a spline to the stream-track in $\phi_1 - V_{GSR}$ coordinates using a Gaussian mixture modelling approach, accounting for contamination by a background population, following the approach described in \citet{koposov_2023_OC}. Rejecting one star with [Fe/H]$> -1.5$ and one more with [Fe/H]$<-3.5$ brings the total number of DESI observations of GD-1 members to  {151 of which 115 are {\it new} } members and for the remaining  {36 stars}, DESI has decreased the radial velocity uncertainty. Slightly relaxing the cuts we made in proper motion, color-magnitude, and the choice of the stellar isochrone used, can all increase the numbers of stars by ~10-30 stars depending on the specific choices made. We consider our selection to be conservative. 

With these new data we spectroscopically identify a broad cocoon structure surrounding a thinner and kinematically colder stream in the region $-30^\circ \leq \phi_1 \leq 2^\circ$ of the GD-1 tidal stream (where the majority of our observations were obtained). This cocoon was first identified by \citet{Malhan_2019_cocoon} using primarily Gaia positions and proper motions and was recently confirmed by \citet{Ibata_2023}. 

Using a bivariate Bayesian mixture modeling algorithm and assuming that the thin stream and cocoon are described by Gaussians, we determine their angular widths to be 
$\Sigma_{\delta\phi_2,\mathrm{thin}} = 0.150\pm0.014^\circ$ and 
 {$\Sigma_{\delta\phi_2,\mathrm{cocoon}} = 1.252\pm0.121^\circ$.} Since our DESI fields are restricted to $|\delta_{\phi_2}| \sim \pm 3^\circ$ from the stream track this could be an under-estimate of the width of the cocoon.
The velocity dispersions of these two components are  {$\Sigma_{V_{GSR},\mathrm{thin}} = 3.09\pm0.76$\kms\ and 
$\Sigma_{V_{GSR},\mathrm{cocoon}} = 7.81\pm1.01$\kms. }
An analysis of the velocity dispersions of the thin stream and cocoon components defined by membership probability using  $\delta\phi_2$ only, but allowing for contamination by a background population gives a nearly identical velocity dispersion for the thin stream component but a somewhat lower velocity dispersion of  {$\sim 4.84$\kms\ } for the cocoon component.

A bivariate analysis also including a background would benefit from more data and is deferred to a future work.

DESI spectroscopy yields  mean [Fe/H] values of   $\left<\mathrm{[Fe/H]_{thin}}\right>$ 
$= -2.54\pm0.04$~dex and 
$\left<\mathrm{[Fe/H]_\mathrm{cocoon}}\right>$ $= -2.47\pm0.06$~dex  respectively, with
intrinsic metallicity dispersions  $\sigma_\mathrm{[Fe/H],thin} = 0.27\pm0.03$~dex and
$\sigma_\mathrm{[Fe/H],\mathrm{cocoon}} = 0.31\pm0.04$dex respectively. 
Our measurements of the mean metallicity  are consistent with previous estimates of [Fe/H] for the whole GD-1  stream and also imply that the cocoon and the thin stream probably originated from the same progenitor. The measured metallicity dispersions are likely overestimated and will be improved.

\subsection{Origin of the ``cocoon'' \label{sec:origin_cocoon}}
While its low metallicity,  retrograde orbit and the thinness and coldness of the most visible part of the stream clearly imply that the progenitor of the GD-1 stream was an accreted GC, the origin of the cocoon is currently uncertain.  At least four possibilities exist.
\begin{enumerate}
\item{The cocoon could comprise GC stars that were tidally stripped from the GC while it was still inside its parent dwarf galaxy (pre-accretion tidal stripping.)} 
\item{It could be a tidal stream from a very metal poor dwarf galaxy, while the thin component could be the stream from an associated nuclear star cluster or GC.} 
\item{A thin tidal stream generated from the GC progenitor could have been heated by multiple interactions with dark subhalos in the MW.}
\item{A thin tidal stream generated from the GC progenitor could have been perturbed and heated by a massive Sagittarius dwarf galaxy.}
\end{enumerate}
We now describe each of these possible origins in turn and discuss how one might distinguish between them with future spectroscopic data.

\paragraph{Cocoon stars originated from pre-accretion tidal stripping of the progenitor GC:}  Simulations   show that star clusters  can experience the strongest tidal force in the first few hundred Myr after formation \citep{Xi_Gnedin_2022}. Once the parent dwarf galaxy is accreted onto the Milky Way and is tidally disrupted, the pre-accretion tidal debris forms the cocoon while the intact GC is also liberated from the parent dwarf galaxy and continues to be tidally stripped forming the thinner GC stream \citep{Carlberg_2018, Malhan_2019_cocoon, Malhan_2020, Carlberg_alger_2023}. 

The structure  and kinematics of cocoons so formed may  be important probes of the central dark matter distribution of the parent dwarf galaxy. \citet{Malhan_2020} carried out a suite of over 100 controlled N-body simulations of dwarf galaxies and their GCs falling into a Milky Way-like potential. They showed that accreted GCs born in dwarf galaxies with cored dark matter halos exhibit fewer morphological peculiarities (e.g., gaps and off-track stream structures like the spur and blob) than their counterparts born in dwarf galaxies with cuspy dark matter halos. They argue that for a parent halo of a given mass, the physical width, radial and tangential velocity dispersions and spread in angular momentum averaged along an accreted GC stream can serve as diagnostics of whether the parent host galaxy had a cusp or a core.  \citet{Malhan_2020} showed that if considering dwarf galaxies of halo mass $10^9$\msun, the gentler tidal fields in a cored dark matter halo would result in a tidal stream with  a velocity dispersion of  $\sim2-5$\kms, while a dwarf galaxy with a cuspy halo would produce a stream velocity dispersion of order 8-12\kms.
In this pre-accretion tidal stripping scenario, a cocoon velocity dispersion of  $\sim 8$\kms\ would imply that GD-1's parent dwarf galaxy had a cuspy dark matter halo while $\sim 5$\kms\ would imply the parent dwarf had a cored halo. Additional observations along a larger stretch of the stream are required to draw definitive conclusions since this formation scenario predicts that cocoon  are not restricted to short segments of the stream.  

\paragraph{Cocoon stars originated from the parent dwarf galaxy:} A second possibility is that the cocoon comprises the stars of GD-1's parent dwarf galaxy. The velocity dispersion of the cocoon as well as its metallicity are reasonably consistent with those of other Local Group dwarf galaxies. Dwarf galaxies tend to show greater metallicity dispersions than GCs but the metallicity distributions of the thin stream and cocoon components are currently statistically indistinguishable, although improved [Fe/H] measurements are needed to make a definitive statement.
Nuclear star clusters  do show large metallicity dispersions \citep[e.g.,][]{Alfaro-Cuello_2019,Nogueras-Lara_2022}, but all known nuclear star clusters are more metal rich than GD-1.

GD-1's low metallicity and retrograde orbit are strong indicators that  its progenitor was accreted from a dwarf galaxy. Previous work based on the location of GD-1 in action space relative to known GCs and halo stars suggests a possible association with debris from accreted dwarf galaxy(ies) such as `Sequoia/Arjuna/ I'itoi' \citep{Myeong_etal_2019, Bonaca2021,Malhan_2022b}. Accreted GCs which are about 60\% of the GC population \citep{Choksi2018,  Choksi_Gnedin2019, Creadey2019} tend to be on more radial orbits and tend to be found at larger radii \citep{ Chen_Gnedin2022, Chen_Gnedin_2023, Belokurov_Kravtsov_2024}. In contrast, in situ GCs are more metal-rich and lie deeper in the Galactic potential (at more negative energy). GD-1 is peculiar because it is quite deep in the potential but rather than being on a radial orbit it is on a retrograde and nearly tangential (high angular momentum) orbit. 

A possible explanation is that it was brought in by a fairly heavy dwarf galaxy that experienced significant dynamical friction \citep{BT}.
Follow-up high-resolution spectroscopy to obtain abundance data will be particularly useful in determining whether the cocoon has abundance signatures characteristic of GCs or dwarf galaxies.  {For example a sodium-oxygen anti-correlation is seen in high-resolution spectra  of GCs with masses at or above $ \sim 2 \times 10^4$\msun\ \citep{Carretta_etal_2010_Na-OAntiCor, Marino_etal_2011, 
Bragaglia_2017_Na-O_anticor} but not in dwarf galaxies. 
Various estimates of the mass of the GD-1 progenitor exist in the literature: $2\times 10^4$\msun\ \citep{koposov2010}, $5\times 10^4$\msun\ \citep{Gialluca_2021}, $7\times 10^4$\msun\ \citep{Bonaca_2019} and $1\times10^5$ \citep{Price-Whelan_2018}. 
While these values are at the lower limit of the mass range of GCs showing multiple stellar populations and a Na-O anti-correlation, 
one reason to be optimistic that future studies may show such an anti-correlation is that a recent study of high resolution spectra of GD-1 
stars (including some in the spur and blob) has already detected 
variations in C and Mg \citep{Balbinot_etal_2022_GD1_met_var}, 
suggesting the presence of multiple populations in this system.}

\paragraph{Cocoon produced by dark matter subhalo impacts with a thin GC stream:}  The third possible origin for the cocoon is heating by impacts from multiple dark subhalos with masses in the range $10^7-10^8$\msun\ \citep[e.g.,][]{Johnston_2002,Ibata_etal_2002, siegal-gaskins_valluri_2008,Carlberg2012,Carlberg2013,Erkal_2015a,Sanders2016,Banik2018,Bonaca_2019,Banik2021,Carlberg_alger_2023}. Simulations provide insights into the origin of the broad cocoon-like components produced by subhalo impacts. Encounters with small impact parameters between thin GC streams and dark subhalos cause the decrease of the angular momenta of stream stars ahead of the point of closest approach, since stars are pulled towards the perturber, and vice-versa. Stars with lower angular momenta than the progenitor spread to smaller galactocentric radii while stars with higher angular momenta spread to larger radii. The accumulation of sub-halo encounters with time is expected to lead to an increase in velocity dispersion \citep{Ngan2015, Ngan2016} along the stream and may also create gaps which orbital shearing will disperse over time. In addition, orbital precession in the non-spherical potential of a Milky Way-like system causes orbits to diverge on a time-scale of several Gyr \citep{Ibata_etal_2002, siegal-gaskins_valluri_2008}. Since the effects of sub-halo encounters are initially fairly local, only old GC streams (from star clusters accreted long ago) will have had time to experience subhalo heating over significant stretches of the stream. Recent models suggest that the GD-1 stream is between 3-5.5~Gyr old \citep{Gialluca_2021}, suggesting it might not be old enough to have been heated by subhalos along its entire extent. On-going and future observations of the GD-1 stream with DESI will enable us to determine whether the cocoon is a local or a much more extended feature.

\paragraph{Perturbations from a massive Sagittarius dwarf:}

\citet{deBoer2020} proposed that the spur in GD-1 could have been caused by interaction of the Sagittarius dwarf spheroidal galaxy (Sgr) with the GD-1 stream.   \citet{Dillamore_etal_2022} showed that if the Sgr dwarf  was originally more massive than $4\times 10^{10}$\msun, gravitational torques from it would affect tidal streams (like Pal-5 and GD-1) that lie within the orbital pericenter radius of Sgr.  Such torques can cause asymmetries between the leading and trailing arms due to the folding back of one of the arms as well as the appearance of off-stream structures like the GD-1 ``spur'' and ``blob'' \citep{Price-Whelan_2018} (the ``blob''  overlaps with the cocoon we characterized in previous subsections). Perturbations from a massive Sgr may even produce structures that appear like a separate stream on the sky but with kinematics similar to the main stream. \citet{Dillamore_etal_2022} show that if Sgr has affected the GD-1 stream it could have also  produced a thin off-set stream similar to Kshir, a stream intersecting GD-1 discovered by \citet{Malhan_2019_Kshir} . 

\medskip

In practice more than one of the above processes could have affected the GD-1 stream to varying degrees: the cocoon could have contributions from both pre-accretion tidal heating, stars from the parent dwarf galaxy and heating from sub-halo encounters. Additional observations, as well as modeling, will help to untangle the contributions from  various processes.

Recently \citet{Ibata_2023} also identified a two-component structure for the GD-1 stream. They used the VLT/UVES spectrograph to follow up potential GD-1 candidates detected in the Gaia EDR3 and DR3 catalogs. The VLT/UVES data have velocity precision of 1\kms. They also obtained observations with the IDS long-slit spectrograph on the 2.5m Isaac Newton Telescope  which have a lower velocity precision ($\sim10$\kms). Using 323 stars with line of sight velocity measurements they find a two component fit with velocity dispersions for the ``thin stream'' to be $\sigma =7.4\pm1.1$\kms\ while the cocoon  has a velocity dispersion $\sigma= 29.1\pm 6.1$\kms\ with a fraction of $0.73\pm0.09$ in the thin  component.  

Our analysis in section~\ref{sec:cocoon} is restricted to a narrow range of $\phi_1$ values and implies a much smaller velocity dispersion of 5-8\kms\ than found by \citet{Ibata_2023}. Their analysis is applied to the entire stream ($-100^\circ <\phi_1< 20^\circ$) while our analysis above is restricted to $-30^\circ <\phi_1< 0^\circ$. Although we have few DESI observations at $\phi_1<-50^\circ$, when we use data from DESI, SDSS+LAMOST, MMT (all the data within the dashed red curves in Figure~\ref{fig:rv_phi1_all}) to estimate the width of the thin stream and cocoon we find that the velocity dispersions of the thin component and cocoon are $1.89\pm0.38$\kms\  and  $7.68\pm0.72$\kms\ respectively with a thin stream fraction $Q= 0.58$.  Although these values are probably dominated by our measurements in the cocoon area,  they are significantly smaller than the estimates from \citet{Ibata_2023}. While our estimated velocity dispersion for the thin stream component is more consistent with previous work on the GD-1 stream's thin component as well as other GC streams. Our estimated velocity dispersion for the cocoon is also more consistent with simulation predictions  \citep{Malhan_2020, Carlberg_alger_2023}. 

At the present time it is not possible to draw definitive conclusions as to the origin of the cocoon component of the  GD-1 stream. The more important conclusion is that the GD-1 stream does have narrow and wide components approximately as expected in Milky Way simulations with an accreting sub-halo population. These initial data suggest that definitive conclusions will be possible as increased numbers of stars in individual streams become available and more streams are studied.

\subsection{Other factors affecting the structure of the GD-1 stream \label{sec:other_effects}}

In addition to the mechanisms discussed in the previous section, there have been recent studies of at least a few other factors that can affect the structure and kinematics of tidal streams like GD-1. 

It is thought that the massive merger in the Milky Way's history referred to as the Gaia-Enceladus-Sausage event \citep{Belokurov_GES_2018, Helmi_GES_2018} could have resulted in the disk of the Milky Way becoming misaligned with the resulting triaxial dark matter halo \citep{Dillamore2022_tilting, Han_2022, Han_2023} or otherwise cause torques on the stellar disk which cause it to tilt as it realigns with the halo. Simulations show that the tilting rate of disks following mergers can have angular speeds of $\sim 10-20$\kms kpc$^{-1}$ \citep[e.g.,][]{Earp_2017, Dodge_2023}. 
Recently \citet{Nibauer_etal_2023} showed that the gravitational torque from such a tilting disk can cause both narrowing and broadening  (depending on the magnitude and direction of tilting) of tidal streams with pericenters within $\sim 20\,$kpc. 
Tilting can also cause changes in the line-of-sight width $\delta \phi_2$ as well as the width of the radial velocity distribution. They show that several of the features of the Pal-5 stream are consistent with a disk tilting rate of 15\kms kpc$^{-1}$. While these authors do not discuss the GD-1 stream it could also be affected by disk tilting since  its  pericenter radius is $<20$~kpc. 

Despite the large number of possible origins for the cocoon described in this and the previous subsection, detailed models and additional spectroscopic data may enable us to determine its origin.

\subsection{GD-1 data expected from future DESI data releases}

Future DESI data releases will include many more GD-1 members. 
 {The first 3 years of the survey have already yielded over 600 spectroscopically confirmed members} in the stream area (2000 square degrees). Targets were selected by CMD and proper motion cuts and most were observed as part of secondary and tertiary observing programs designed for future survey planning purposes.  The regions with confirmed members span $-75^\circ <\phi_1 <15^\circ$.  The sample of confirmed members include stars down to $r=21$ (the velocity uncertainty at the faint end is $\sim 15$\kms). Current plans for the future include observing up to ninety 3$^\circ$ diameter fields in the GD-1 stream region. It is anticipated that by the end of the 5-year program DESI will obtain spectra for $\sim 3000$ GD-1 stream members, making it one of the most thoroughly surveyed GC streams.

\acknowledgements
MV gratefully acknowledges support from NASA-ATP grants 80NSSC20K0509  and 80NSSC24K0938. SK acknowledges support from the Science \& Technology Facilities Council (STFC) grant ST/Y001001/1. 
T.S.L. acknowledges financial support from Natural Sciences and Engineering Research Council of Canada (NSERC) through grant RGPIN-2022-04794. 
APC acknowledges support from a Taiwan Ministry of Education Yushan Fellowship and the Taiwan National Science and Technology Council (NSTC) grants 109-2112-M-007-011-MY3 and 112-2112-M-007-017-MY3.
This material is based upon work supported by the U.S. Department of Energy (DOE), Office of Science, Office of High-Energy Physics, under Contract No. DE–AC02–05CH11231, and by the National Energy Research Scientific Computing Center, a DOE Office of Science User Facility under the same contract. Additional support for DESI was provided by the U.S. National Science Foundation (NSF), Division of Astronomical Sciences under Contract No. AST-0950945 to the NSF’s National Optical-Infrared Astronomy Research Laboratory; the Science and Technology Facilities Council of the United Kingdom; the Gordon and Betty Moore Foundation; the Heising-Simons Foundation; the French Alternative Energies and Atomic Energy Commission (CEA); the National Council of Science and Technology of Mexico (CONACYT); the Ministry of Science and Innovation of Spain (MICINN), and by the DESI Member Institutions: \url{https://www.desi.lbl.gov/collaborating-institutions}. Any opinions, findings, and conclusions or recommendations expressed in this material are those of the author(s) and do not necessarily reflect the views of the U. S. National Science Foundation, the U. S. Department of Energy, or any of the listed funding agencies.

The DESI Legacy Imaging Surveys consist of three individual and complementary projects: the Dark Energy Camera Legacy Survey (DECaLS), the Beijing-Arizona Sky Survey (BASS), and the Mayall z-band Legacy Survey (MzLS). DECaLS, BASS and MzLS together include data obtained, respectively, at the Blanco telescope, Cerro Tololo Inter-American Observatory, NSF NOIRLab; the Bok telescope, Steward Observatory, University of Arizona; and the Mayall telescope, Kitt Peak National Observatory, NOIRLab. NOIRLab is operated by the Association of Universities for Research in Astronomy (AURA) under a cooperative agreement with the National Science Foundation. Pipeline processing and analyses of the data were supported by NOIRLab and the Lawrence Berkeley National Laboratory (LBNL). The Legacy Surveys also use data products from the Near-Earth Object Wide-field Infrared Survey Explorer (NEOWISE), a project of the Jet Propulsion Laboratory/California Institute of Technology, funded by the National Aeronautics and Space Administration. Legacy Surveys was supported by: the Director, Office of Science, Office of High Energy Physics of the U.S. Department of Energy; the National Energy Research Scientific Computing Center, a DOE Office of Science User Facility; the U.S. National Science Foundation, Division of Astronomical Sciences; the National Astronomical Observatories of China, the Chinese Academy of Sciences and the Chinese National Natural Science Foundation. LBNL is managed by the Regents of the University of California under contract to the U.S. Department of Energy.

This work has made use of data from the European Space Agency (ESA) mission Gaia\footnote{\url{https://www.cosmos.esa.int/gaia}}, processed by the Gaia Data Processing and Analysis Consortium (DPAC\footnote{\url{https://www.cosmos.esa.int/web/gaia/ dpac/consortium}}).
This work has made use of data from the Sloan Digital Sky Survey. Funding for SDSS-III has been provided by the Alfred P. Sloan Foundation, the Participating Institutions, the National Science Foundation, and the U.S. Department of Energy Office of Science. The SDSS-III web site is \url{http://www.sdss3.org/}. This work also utilized data from the LAMOST survey. 

This paper made use of the Whole Sky Database (wsdb) created and maintained by Sergey Koposov at the Institute of Astronomy, Cambridge, with financial support from STFC and the European Research Council (ERC).
\medskip

The authors are honored to be permitted to conduct scientific research on Iolkam Du’ag (Kitt Peak), a mountain with particular significance to the Tohono O’odham Nation. The University of Michigan is located on the traditional territory of the Anishinaabe people. We acknowledge, their contemporary and ancestral ties to the land and their contributions to the university.

For the purpose of open access, the author has applied a Creative Commons Attribution (CC BY) licence to any Author Accepted Manuscript version arising from this submission.

\software{astropy \citep{astropy2018, astropy2022}, 
emcee \citep{emcee},
gala \citep{gala2017,Gala_APW},
matplotlib \citep{matplotlib},
numpy \citep{harris_array_2020}, 
scipy \citep{virtanen_scipy_2020},
scikit-learn \citep{scikit-learn},       
stan \citep{Carpenter_etal_2017}}.

\facilities{KPNO: Mayall (DESI), Gaia, Blanco (DECam).}
\medskip

{\it \large Data Availability:} {Data behind all figures in this paper are available as FITS files at: \url{https://zenodo.org/records/11638329}}  DOI: 10.5281/zenodo.11638329.

\appendix 
\section{Measuring the distance stream track \label{sec:distance}}

\begin{figure*}[ht]
    \begin{centering}
    \includegraphics[trim=0 0 0 0, clip,width=0.378\textwidth]{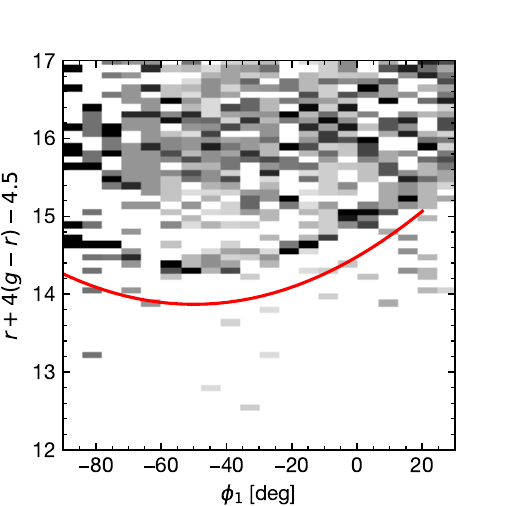}
    \includegraphics[trim=0 10 0 0, clip ]{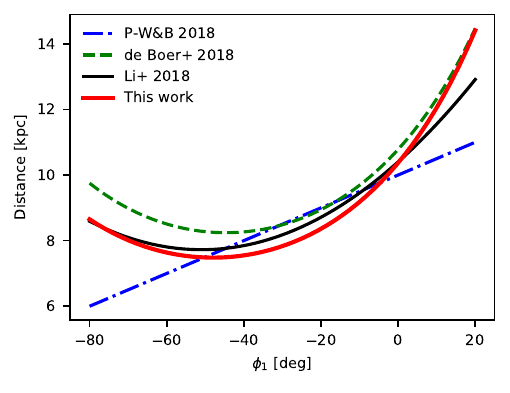}
    \caption{Left: The stream distance gradient from the subgiant branch stars. The greyscale shows the distribution of the $r+4(g-r)-4.5$ (a proxy for distance modulus) as a function of $\phi_1$ for stars with $.25<g-r<.4$. The subgiant branch stars form a narrow sequence in the figure. The red curve shows the adopted distance gradient model (shifted down by 0.5 magnitudes for clarity).
    Right: Our estimated distance to the stream as a function of $\phi_1$ compared with relations from the literature.  }
    \end{centering}
    \label{fig:distance_gradient}
\end{figure*}

To calibrate the distance gradient along the stream, we use the proper motion and parallax selection described in Section~\ref{sec:membership} and then focus on the subgiant branch stars as those form a narrow sequence on the CMD (see Figure~\ref{fig:cmd}). Specifically we select stars with $0.25<g-r<0.4$ and assume that the subgiant branch has an approximately constant $r + 4 (g-r)$. 
Figure~\ref{fig:distance_gradient} (left) shows the $r+4(g-r)-4.5$ vs. $\phi_1$ for likely GD-1 subgiant stars. The red curve shows our fiducial distance modulus vs. $\phi_1$  relation. Figure~\ref{fig:distance_gradient} (right) shows various relations for the distance to the GD-1 stream as a function of $\phi_1$ from \citet{Price-Whelan_2018} (blue dot-dashed), \citet{deBoer2018} (green dashed), \citet{LiYannyWu2018} (black solid)  and equation~1 (this work, red solid).

\section{Table of DESI-EDR measurements in the GD-1 area \label{sec:datatable}}

Table~4 gives DESI-EDR measurements of $V_{GSR}$ and [Fe/H] and their 1$\sigma$ measurement uncertainities, $V_{err}$ and [Fe/H]$_{err}$ are from the RVS pipeline. 
The $V_{GSR}$ measurements have been corrected by 0.93\kms, the systematic error on RV measurements reported in \citep{koposov_edr_2024}. 
No corrections have been applied to [Fe/H] measurements. {\tt TARGET\_ID} is the DESI target ID, RA and Dec values are from  Gaia-DR3.

\begin{table}
\begin{tabular}{ccccccccc}
\multicolumn{9}{c}{{\bf Table 4: DESI-EDR Observations of GD-1 region (data in Fig~4a) }}\\
\hline
TARGET-ID &  RA &  Dec & $\phi_1$  & $\phi_2$  & $V_{GSR}$  & $V_{err}$   & [Fe/H]  & [Fe/H]$_{err}$ \\
     & [deg]& [deg]& [deg]& [deg]& [\kms] & [\kms] & [dex] & [dex] \\
\hline
39633374\ldots & 192.76\ldots & 58.91\ldots & -3.70\ldots & 0.49\ldots & 125.44\ldots & 1.91\ldots & -1.92\ldots & 0.05 \\
39633362\ldots & 191.93\ldots & 57.92\ldots & -4.44\ldots & -0.29\ldots & 113.64\ldots& 1.49\ldots & -1.50\ldots & 0.04 \\
39633377\ldots & 191.69\ldots & 59.30\ldots & -4.10\ldots& 1.03\ldots & 130.81\ldots & 1.74\ldots & -1.75\ldots & 0.04 \\
\ldots & \ldots & \ldots & \ldots & \ldots & \ldots & \ldots & \ldots & \ldots \\
\ldots & \ldots & \ldots & \ldots & \ldots & \ldots & \ldots & \ldots & \ldots \\
\end{tabular}
\end{table}

\newpage
\bibliography{gd1_paper}{}
\bibliographystyle{aasjournal}
\end{document}